\documentclass[preprint,showpacs,preprintnumbers,amsmath,amssymb]{revtex4}

\usepackage{graphicx}
\usepackage{dcolumn}
\usepackage{bm}

\newcommand\beqa{\begin{eqnarray}}
\newcommand\eeqa{\end{eqnarray}}
\newcommand\n{\nonumber\\}

\begin{document}
\allowdisplaybreaks{

\preprint{KEK-TH-1586}

\vspace*{5mm}

\title{Family Unification via \\Quasi-Nambu-Goldstone Fermions 
in String Theory}

\author{Shun'ya Mizoguchi}
\altaffiliation[Also at ]{Department of Particle and Nuclear Physics, The Graduate University for Advanced Studies}
\email{mizoguch@post.kek.jp}
\affiliation{%
High Energy 
Accelerator Research Organization (KEK)\\
Tsukuba, Ibaraki 305-0801, Japan 
}
%
\author{Masaya Yata}
\altaffiliation[ ]{JSPS fellow}
\email{yata@post.kek.jp}
\affiliation{%
The Graduate University for Advanced Studies\\
Tsukuba, Ibaraki 305-0801, Japan 
}

\date{Nov.26, 2012\\ \phantom{} }

\begin{abstract}
Some of the supersymmetric nonlinear sigma models on exceptional groups 
($E_7$ or $E_8$) are known to yield almost minimal necessary content 
of matter fields for phenomenological model building, including three
chiral families and a Higgs multiplet. We explore a possible 
realization of such a family unification in heterotic string theory, 
where the 
spontaneously 
broken symmetry
that we focus on   
is 
the one associated with the change of global orientation of 
the spin bundle embedded in the gauge bundle,
which cannot be ignored 
in the presence of a magnetic source of the $B$ field. 
We show in a simple model that
it indeed gives rise to chiral matter fields on the 
defect similarly to the domain-wall fermions, and  
compare the chiral zero mode spectrum with the sigma model. 
The setup is reduced to a system similar to that postulated in the orbifold GUT or the 
grand gauge-Higgs unification model, where the origin of ${\bf Z}_2$ identification 
is accounted for as arising from the bolt of the Atiyah-Hitchin manifold.

\end{abstract}

\maketitle

\section{\label{sec:Introduction}Introduction
}
One of the long-standing questions in particle physics is why there are 
three families of quarks and leptons in Nature. 
While grand unification successfully unifies gauge 
interactions and gathers quarks and leptons in one generation 
into an irreducible representation, the repetitive structure of fermions  
with identical quantum numbers has 
been a mystery for decades.

For almost thirty years superstring theory 
has been long thought of as 
a promising framework for   
describing the 
ultra-violet dynamics of all interactions in a unified way,
yet it is still not clear 
precisely how the standard model is realized in it.
One of the major obstacles in constructing realistic phenomenological 
models in string theory is the appearance of numerous moduli, which are
ever present in any smooth supersymmetric compactifications.
Roughly speaking, moduli are  parameters 
of the internal six-dimensional manifold and any structures thereof, 
and
give rise to massless 
scalars in the low-energy field theory.
They are problematic for phenomenological applications of string theory 
for 
the following reasons:

First, they may cause the cosmological moduli problem \cite{cosmo_moduli}.
If they are  
present to date, their masses
must be light enough in order for their energy density
not to exceed the observed cosmological bound.    
On the other hand, if they are assumed to have decayed, there is a risk that they 
may spoil the success of the BBN scenario
or the released entropy may dilute the baryon asymmetry produced by that time, 
depending on the time they decay.
These problems might be avoided by modifying the conventional 
cosmological scenario.

The second reason concerns the moduli stabilization.
There are many ways to realize standard-like models in string theory;
each option has advantages and drawbacks. 
Moduli stabilization is the most successful in type IIB stirng theory, where 
the KKLT \cite{KKLT} and LARGE volume scenarios \cite{LVS} are well known. 
However, it was pointed out that \cite{tension}  
to combine the moduli stabilization 
sector and the standard model sector is sometimes not straightforward. 
The D-brane models \cite{Dbranemodels} also 
have a difficulty 
that the up-type Yukawa couplings can be generated only nonperturbatively 
\cite{BKLO,BCLRW}. 
On the other hand, the model building based on $E_8 \times E_8$ heterotic sting is 
the oldest and still attractive 
approach, 
but to completely fix all the moduli is harder 
than type II theories \cite{heteroticmoduli,heteroticmoduli2,heteroticmoduli3,heteroticmoduli4,
heteroticmoduli5}.

Finally, last but not least, the numerous moduli 
lead to many different possible string vacua - the landscape -
which is also due to the existence of numerous 
possible  
compactification 
manifolds. This is a problem since it prevents us from uniquely 
determining the high-energy extension of the standard model.
None of the string theory realizations of a standard-like model known  
to date do not 
{\em explain} why we observe three families.  
Although the anthropic principle is indeed a constraint from the 
undeniable ``observation" that 
we %
live in the Universe, %
it
would not be strong enough to single out, for instance, 
the correct type of Calabi-Yau (if any) or the precise values of fluxes 
from a huge number of possibiities.
The statistical approach \cite{statistical} has shown no evidence of 
probabilistic dominance of standard-like models. On the contrary, 
standard-like models are rather rare \cite{rare}.

In field theory (as opposed to string theory), 
an interesting proposal had been put forward in \cite{BPY}
long time ago, even before superstring theories have come 
to be known,
where it was suggested that 
the quarks and leptons %
could be understood as quasi-Nambu-Goldstone %
fermions \cite{BLPY}
of a supersymmetric nonlinear sigma model \cite{BPY,Ong,KY,IY,BKMU,IKK}.
The idea 
that 
the family structure comes from a group theory
is an old one \cite{old_Family_unification} 
\footnote{However,
most of the early family unification models 
suffered from the problem of mirror families. See  
\cite{0202178} for %
more %
discussion.}.
The advantage of the coset sigma model approach 
is that the associated quasi-Nambu-Goldstone fermions are typically 
chiral. 
Remarkably enough, it was shown by Kugo and Yanagida \cite{KY}
that the 
sigma model based on $E_7/(SU(5)\times SU(3)\times U(1))$ 
yields precisely {\em three} sets of chiral superfields transforming in 
${\bf 10}\oplus {\bf\bar 5}$ in addition to a single ${\bf 5}$ of $SU(5)$ 
as the target space, the former of which may be identified as three generations 
of quark and lepton supermultiplets, and the latter a Higgs multiplet, 
respectively. 
Thus, the Kugo-Yanagida model offers 
a chance to realize 
almost minimal necessary matter content 
for an $SU(5)$ GUT
in an amazingly economical way\footnote{For its 
anomalies see section \ref{summary}.}.

In this paper  
we propose a  
scenario for 
realizing  
such a family unification in {\em heterotic string theory} \cite{heterotic_string}.
The spontaneously broken symmetry that we focus on is 
the one associated with the change of global orientation of 
the spin bundle embedded in the gauge bundle,
which becomes relevant \cite{KM,KM2}
in the presence of a magnetic source of the 
$B$ field, that is, the NS5-branes \cite{SJR,CHS}. 
It has been known for some time that the symmetric 5-brane
in heterotic string theory supports charged chiral fermions \cite{CHS,KM2},
which can be regarded as the quasi-NG fermions associated with 
the spontaneously chosen gauge field configuration of the brane.
We consider a system of two stacks of intersecting 5-branes in the 
$E_8 \times E_8$ theory, which is expected to realize the 
K\"{a}hler coset $E_8/(E_6\times H)$ for some subgroup $H$ containing a 
$U(1)$ factor.  We will perform an explicit computation 
of the Dirac zero modes on a compactified smeared intersecting 
5-brane background to find two chiral and one anti-chiral zero modes 
in the {\bf 27} representation of $E_6$. This agrees with the chiral 
spectrum of the supersymmetric $E_8/(E_6\times H)$ sigma model, 
confirming the expectation.

For the purpose of obtaining more realistic cosets, 
we also perform a similar computation in the background with 
a constant $U(1)$ Wilson line 
in addition to the $SU(3)$ gauge configuration.
This time, unfortunately, 
a pair of chiral and anti-chiral modes turn out to appear and 
are asymmetrically localized on different branes, 
being unsatisfactory as a realization of the $E_8/(SO(10)\times H)$ 
coset.
We will speculate about the non-standardly embedded NS5-branes 
that may lead to 
realistic K\"{a}hler cosets having three 
generations 
(with a possible anti-chiral generation and or a pair of  
chiral and anti-chiral generations).

The idea that chiral fermions are localized on a defect is 
a familiar one (e.g. \cite{CHinflow,Kaplan}), 
so it is not a crazy thought.
On the other hand, this setup is quite different from the conventional 
smooth Calabi-Yau compactifications.
It is somewhat similar to the old setup by Witten \cite{WittenO(32)}, but the crucial 
difference is that our setup contains NS5-branes.
Of course, NS5-branes are difficult objects to deal with; 
the supergravity solution for NS5-branes
develops an infinite throat, and the dilaton blows up as it goes down 
the throat.  
The microscopic description is less  
understood than D-branes. 
However, the idea is that one might capture their {\em low-energy} 
dynamics by a nonlinear sigma model, just in the same way as one can
describe pions even if one did not know anything about QCD.

The organization of this paper is as follows. In section \ref{section2}, we 
give a brief review of the supersymmetric nonlinear sigma models 
with a target space being a K\"{a}hler coset of some exceptional 
Lie group. In section \ref{section3}, we turn to the construction of an 
NS5-brane system that may realize the 
symmetry breaking.
We first show what symmetry in heterotic string theory is spontaneously 
broken in the presence of NS5-branes. We then solve explicitly the 
gaugino Dirac equation on a compactified smeared background and compare 
the chiral zero mode spectrum with that of the corresponding 
sigma model. 
The setup is reduced to a system similar to that postulated in the 
orbifold GUT or the
(grand) gauge-Higgs unification.
We also suggest a possible interpretation of 
the negative tension branes in heterotic string theory as a superposition 
of T-duals of the Atiyah-Hitchin manifold. 
In section \ref{Wilson_line}, a similar computation is performed 
with a constant Wilson line included in one of the transverse directions.   
We conclude in section \ref{summary} with a summary 
and discussion on possible future directions.
Some useful facts about $E_8$ and conventions for the gamma 
matrices are collected in two appendices.

\section{Exceptional coset supersymmetric nonlinear sigma models}
\label{section2}
A nonlinear sigma model on a group coset 
$G/H$ describes the low-energy dynamics of Nambu-Goldstone (NG) bosons 
which arise when a global symmetry $G$ is spontaneously broken to  
its subgroup $H$. In supersymmetric nonlinear sigma models those 
NG bosons are accompanied by their super-partners called 
quasi-Nambu-Goldstone (qNG) fermions. In four dimensions the target 
space of an ${\cal N}=1$ nonlinear sigma model is a K\"{a}hler 
manifold, and the chiralities of the qNG fermions can readily be 
determined by examining their ``$Y$-charges" \cite{IKK} of the K\"{a}hler 
coset, as we briefly review below.

Let $G$ be a compact, semi-simple group, and $H$ be its subgroup. 
The classic theorem of Borel asserts that the coset space $G/H$ is 
K\"{a}hler if and only if $H$ is a centralizer of some torus subgroup of $G$, 
that is, $H$ is the group consisting of all elements that commute with  
some $U(1)^n$ subgroup of $G$. This means that if $G/H$ is K\"{a}hler, 
any element in $G/H$ has nonzero charge for some $U(1)$ subgroup. 
Therefore, one can take some linear combination of the $U(1)$ charges so that 
the complexified Lie algebra of $G/H$ is decomposed into a direct sum of 
positive- and negative-charge eigenspaces. We fix some such a combination 
and call it ``$Y$-charge", following \cite{IKK}. Let $X^I$ ($I=1,\ldots,k$) be generators 
having negative $Y$-charge, with $k$ being half 
the real dimensions of $G/H$, and consider the ``BKMU variable"
\beqa
\xi(\phi)&\equiv&e^{\phi^I X^I},
\eeqa
where $\phi^I$'s are the set of chiral superfields parameterizing the K\"{a}hler coset.
Then it was shown that the  K\"{a}hler potential can be 
expressed in terms of $\xi(\phi)$ \cite{BKMU,IKK} (see also \cite{AAvH,DV} for earlier 
discussions), and a certain $H$-invariant projection operator 
$\eta$ acting on the representation vector space. 
Moreover, the correspondence between the $Y$-charge and the complex 
structure of the K\"{a}hler manifold is one-to-one. Therefore, in order to 
determine the chiral spectrum of a given coset, we have only to choose 
a suitable combination of $U(1)$ charges as the $Y$-charge and
examine which generators 
have negative $Y$-charges.

We now turn to the actual decompositions in the relevant examples:
\subsection{$E_7/( SU(5) \times SU(3) \times U(1))$}
This is the original coset space considered by Kugo and Yanagida \cite{KY}. The adjoint ${\bf 133}$ 
of $E_7$ is decomposed into a sum of 
irreducible representations of $SU(5) \times SU(3)$ as follows:
\beqa
{\bf 133}&=&
({\bf 24},{\bf 1})_0\oplus
({\bf 1},{\bf 8})_0\oplus
({\bf 1},{\bf 1})_0\oplus
({\bf 5},{\bf \bar 3})_{4}\oplus
({\bf \bar 5},{\bf 3})_{-4}
\n
&&\oplus 
({\bf 5},{\bf 1})_{-6}\oplus
({\bf \bar 5},{\bf 1})_{6}\oplus
({\bf 10},{\bf \bar 3})_{-2}\oplus
({\bf \overline{10}},{\bf 3})_{2},
\label{E_7decomposition}
\eeqa
where the subscript numbers indicate the $U(1)$ charges of 
the representations they follow. 
The charge normalization is taken so that it coincides with $-h_\sharp$ 
(\ref{h_sharp}). (The minus sign is needed to agree with the conventions 
of the chirality in the literature.)  
The representations in the first line form 
an $SU(8)$ subalgebra, whereas those in the second line do a rank-4 
antisymmetric tensor representation ${\bf 70}$ of $SU(8)$, the familiar 
realization of $E_7$ in supergravity \cite{CJ}. Those which have negative 
charges are
\beqa
({\bf \bar 5},{\bf 3})_{-4},({\bf 10},{\bf \bar 3})_{-2},  ({\bf 5},{\bf 1})_{-6}.
\label{KYspectrum}
\eeqa
Therefore, the qNG fermions of this model are three sets of ${\bf 10}\oplus{\bf \bar 5}$ 
 and a single ${\bf 5}$ of $SU(5)$.

\subsection{$E_8/( SU(5) \times SU(3) \times U(1)^2)$}
It has also been known for some time that the net number of chiral 
qNG fermions does not change if the coset group is extended from 
$E_7$ to $E_8$ with an appropriate choice of the $U(1)$ $Y$ charge.
The relevant decompositions are 
\beqa
E_8&\supset& E_7 \times SU(2) %
 \n
{\bf 248}&=&({\bf 133},{\bf 1})\oplus ({\bf 56},{\bf 2})\oplus ({\bf 1},{\bf 3})
\label{E8/E7}
\eeqa
\beqa
E_7&\supset& SU(5) \times SU(3) \times U(1)_{-h_\sharp} \n
{\bf 56}&=&
({\bf 5},{\bf  3})_{-1}\oplus
({\bf \bar 5},{\bf \bar 3})_{1}
\oplus
({\bf 1},{\bf 3})_{5}\oplus
({\bf 1},{\bf \bar 3})_{-5}\oplus
({\bf 10},{\bf 1})_{3}\oplus
({\bf \overline{10}},{\bf 1})_{-3},
\label{56decomposition}
\eeqa
where the $U(1)$ factor contained in $E_7$ is specifically called 
$U(1)_{-h_\sharp}$ so as to distinguish from the one coming from 
$SU(2)$. The decomposition of $({\bf 133},{\bf 1})$ is the same as 
(\ref{E_7decomposition}). The $E_7$ singlet $({\bf 1},{\bf 3})$ %
has $h_\sharp$ charge $0$. 

If $U(1)_{h_\sharp}$ is again chosen to be the $Y$ charge, 
then in addition to (\ref{KYspectrum})
the new qNG fermions 
$({\bf 5},{\bf  3})_{-1}$, $({\bf 1},{\bf \bar 3})_{-5}$ and $({\bf \overline{10}},{\bf 1})_{-3}$
(which are also doublets of $SU(2)$) arise, 
destroying the excellent similarity 
to the pattern of the observed chiral fermions.
However, let  $h_8$ be the
generator of 
the $U(1)$ subalgebra of the $SU(2)$ such that a pair in the doublet ${\bf 2}$ 
 have eigenvalues $\pm 1$ of opposite signs,
defining
\beqa
Y=-h_\sharp + c~ h_8
\eeqa
with a real constant $c$. 
Then, for example,  if $c$ is taken to be $4$, 
then (\ref{56decomposition}) is replaced by
\beqa
E_8&\supset& SU(5) \times SU(3) \times U(1)_Y \n
({\bf 56},{\bf 2})&=&
({\bf 5},{\bf  3})_{-1\pm4}\oplus
({\bf \bar 5},{\bf \bar 3})_{1\pm4}
\oplus
({\bf 1},{\bf 3})_{5\pm4}\oplus
({\bf 1},{\bf \bar 3})_{-5\pm4}\oplus
({\bf 10},{\bf 1})_{3\pm4}\oplus
({\bf \overline{10}},{\bf 1})_{-3\pm4},
\label{562decomposition}
\eeqa
where the subscripts $\pm$ are understood to mean 
the direct sum of these spaces.
In this way the values of $Y$ split in each doublet.
The negative $Y$-charge components are
\beqa
({\bf 5},{\bf  3})_{-5},~~~
({\bf \bar 5},{\bf \bar 3})_{-3},~~~
({\bf 1},{\bf \bar 3})_{-5\pm4},~~~
({\bf 10},{\bf 1})_{-1},~~~
({\bf \overline{10}},{\bf 1})_{-7},
\eeqa
hence %
nonchiral after the $SU(3)$ breaking. This leaves the same chiral 
spectrum as that of the previous example $E_7/( SU(5) \times SU(3) \times U(1))$.

\subsection{$E_8/( SO(10) \times SU(3) \times U(1))$}
\label{E8/(SO(10)xSU(3)xU(1))}
In this case the decomposition reads 
\beqa
{\bf 248}&=&
({\bf 45,1})_0 \oplus
({\bf 16,1})_3 \oplus
({\bf \overline{16},1})_{-3} \oplus
({\bf 1,1})_0 
\n&&
\oplus
({\bf 16,3})_{-1} \oplus
({\bf 10,3})_2 \oplus
({\bf 1,3})_{-4}
\n&&
\oplus
({\bf \overline{16},\bar 3})_{1} \oplus
({\bf 10,\bar 3})_{-2} \oplus
({\bf 1,\bar 3})_{4}
\n&&
\oplus
({\bf 1,\bar 8})_{0},
\eeqa
where the $U(1)$ charges shown as subscripts are 
those of $3 h_\perp$ (\ref{h_perp}). 
Each line is $E_6$ irreducible. 
One can read off from this decomposition that the negative charge 
components are
\beqa
({\bf \overline{16},1})_{-3},
({\bf 16,3})_{-1},
({\bf 1,3})_{-4},
({\bf 10,\bar 3})_{-2},
\eeqa
which contain three chiral ${\bf 16}$ generations and one ${\bf \overline{16}}$
anti-chiral generation. It has been proved that, even if the $SU(3)$ is further broken 
to $U(1)^2$, one can not obtain four chiral (instead of three chiral plus one anti-chiral) 
fermions for any choice of the $U(1)$ $Y$ charge \cite{IKK}.

\subsection{$E_8/( E_6 \times SU(2) \times U(1))$}
The final example is the coset, which, as we will show in later sections, is to be 
realized by a system of two stacks of intersecting heterotic 5-branes with the 
standard embedding. The decomposition reads
\beqa
{\bf 248}&=&
({\bf 78,1})_0 \oplus
({\bf 27,2})_{-1} \oplus
({\bf 27,1})_2 \oplus
({\bf \overline{27},2})_{1} \oplus
({\bf \overline{27},1})_{-2} \oplus
({\bf 1,1})_0.
\eeqa

\section{Spontaneous symmetry breaking by NS5-branes in heterotic string theory}
\label{section3}
\subsection{What symmetry is broken spontaneously?}
\label{What_symmetry}

In the previous section we have seen that some of 
the supersymmetric nonlinear sigma models on $E_8$ coset 
spaces have attractive chiral matter spectra for particle physics 
model building. In this section we explore the possibility of realizing 
such nonlinear sigma models by using $NS5$-branes in $E_8\times E_8$ 
heterotic string theory.

First of all, if these sigma models are realized in any setup, some symmetry 
must be broken spontaneously. What symmetry is broken spontaneously 
in heterotic string theory? To understand this point, let us recall the well-known 
5-brane solution \cite{SJR,CHS} in heterotic string theory:
\begin{eqnarray}
g_{ij}&=&\eta_{ij}~~~(i,j=0,1,\ldots,5),\nonumber\\
g_{\mu\nu}&=&e^{2\phi}\delta_{\mu\nu}~~~(\mu,\nu=6,\ldots,9),\nonumber\\
e^{2\phi}&=&e^{2\phi_0}+\frac {n\alpha'}{x^2},\nonumber\\
H_{\mu\nu\lambda}&=&-\epsilon_{\mu\nu\lambda}{}^{\rho}\partial_\rho \phi,\n
A_\mu^{\alpha\beta}%
&=&2\rho^2\sigma^{\alpha\beta}{}_{\mu\lambda} \cdot
\frac{x^\lambda}{x^2(x^2+\rho^2)},
\label{heterotic5}
\eeqa
where 
$x^2\equiv \sum_{\mu=6}^9(x^\mu)^2$,
and 
$\epsilon^{\mu\nu\lambda\rho}$ 
is the (undensitized) completely antisymmetric tensor. 
All other components of $H$ vanish.

They satisfy the equations of motion of the low-energy effective supergravity 
of heterotic string theory to leading order in the $\alpha'$ expansion. The bosonic 
part of the Lagrangian is given by
\begin{eqnarray}
{\cal L}&=&\frac1{2\kappa^2} \int d^{10}x  \sqrt{-g}e^{-2\phi}
\left\{
R(\omega)-\frac13 H_{MNP}H^{MNP}+4(\partial_M\phi)^2\right.
\nonumber\\
&& \left. \hspace{40mm}
-\alpha'\left(\frac{1}{30} \text{Tr} (F_{MN}F^{MN})
-R_{MNAB}(\omega_+) R^{MNBA}(\omega_+)\right)
\right\}
\label{Lagrangian}
\end{eqnarray}to this order. 
If all $A_M$ is set to zero in (\ref{heterotic5}), the set of field 
configurations is reduce to the neutral $NS5$-brane solution for 
type II theories.
The gauge field configuration (\ref{heterotic5})
is higher-order in $\alpha'$, and can be obtained by the so-called standard 
embedding
\beqa
A_M^{AB}&=&\omega_M^{~~AB}+H_M^{~~AB}
\eeqa
with $\omega_M^{~~AB}$ being the spin connection.
It is a solution of the well-known heterotic Bianchi identity \cite{HullPLB178(1986)357}
\begin{eqnarray}
dH&=&\alpha'\left(
\text{tr}  R(\omega_+)\wedge R(\omega_+) -\frac1{30} \text{Tr} F\wedge F
\right), 
\label{Bianchi_H}
\end{eqnarray}
whose right hand side is required by the celebrated Green-Schwarz mechanism 
of anomaly cancellation.

Now the point is that the way of embedding the (generalized) spin connection 
$(=\omega+H)$
in $E_8$ is not unique. In the configuration 
(\ref{heterotic5}), %
a particular $SU(2)$ subalgebra of $E_8$ 
is chosen to be set equal to the spin connection, thereby the rotational 
symmetry of the gauge orientation is spontaneously broken.
The number of moduli is that of generators which 
nontrivially act on
the gauge configuration, hence
\beqa
248-133=115,
\eeqa
where $133$ is the dimension of the commutant, $E_7$, of $SU(2)$ in $E_8$.
It was pointed out \cite{CHS} that these 115 moduli, together with 4 translation 
and one scale moduli, in all 120 give rise to massless scalars on the 5-brane 
to form 30 six-dimensional hypermultiplets in heterotic string theory.

The gauge configuration of the symmetric 5-brane (\ref{heterotic5}) has
\beqa
\frac1{480\pi^2}\mbox{Tr} F\wedge F&=&1,
\label{nu}
\eeqa
and hence 
may be regarded as a gauge instanton in the transverse four-dimensional 
space. 
In a flat space, the dimensions of moduli parameters of gauge instantons 
are given by 
\beqa
4 C(G) k-d(G),
\label{moduliformula} 
\eeqa
where $C(G)$ and $d(G)$ are respectively  
the quadratic Casimir and dimension of the compact gauge group $G$,
if the instaton number $k$ is large enough \cite{BCGW}. If $G=E_8$, 
it is valid if $k \geq 3$. Although this formula was derived by using the 
Atiyah-Singer index theorem, the number was also given an interpretation 
\cite{BCGW} as 
that required to label (in addition to the scales, positions and sizes) 
the group orientations of $k$ $SU(2)$ instantons 
embedded in the group $G$. In the $E_8$ case, the Casimir is $30$, 
and the number is $120k-248$, the number $120$ being in agreement with 
the above 5-brane moduli.

However, there is a crucial difference between the moduli of the heterotic 
5-brane and those of gauge instantons in a flat space.
If the instanton number $k$ is one (as in (\ref{nu})), the moduli formula 
(\ref{moduliformula}) is not correct, but a single $E_8$ instanton 
has only $5$ moduli in the four-dimensional Euclidean space \cite{BCGW}. 
This means that the apparent $115$ variations are pure gauge modes and 
do not change physics. On the other hand, unlike the flat-space gauge instantons, 
these gauge variations in heterotic string theory {\em do} change physics by 
leaving their traces on the locations of the magnetic source of the $B$ field, 
that is, the 5-branes.
This is due to %
the %
Green-Schwarz counter term contained 
in the heterotic Lagrangian \cite{KM,KM2}.

According to conventional wisdom in field theory, one might think that 
there would be no such moduli because the NG modes associated with a break-down 
of a {\em local} symmetry are eaten by the gauge fields. This indeed 
also happens here; let %
$\delta_{\Lambda^{E_7}}$, 
$\delta_{\Lambda^{({\bf 56},{\bf 2})}}$ 
and 
$\delta_{\Lambda^{SU(2)}}$ be 
$E_8$ gauge transformations 
corresponding to the decomposition 
(\ref{E8/E7}), 
with gauge parameter functions 
$\Lambda^{E_7}(x^i)$, 
$\Lambda^{({\bf 56},{\bf 2})}(x^i)$ 
and 
$\Lambda^{SU(2)}(x^i)$ 
depending only on the coordinates parallel to the brane, 
then
$\delta_{\Lambda^{({\bf 56},{\bf 2})}}A_\mu$ 
and 
$\delta_{\Lambda^{SU(2)}}A_\mu$
are nonzero and satisfy the linearized equations of motion,
which can be undone by counter gauge transformations on $A_i$.
This is the ordinary Higgs mechanism, where the gauge components 
in the transverse dimensions play the role of adjoint Higgs fields, 
consequently the gauge symmetry is broken to $E_7$. The 
gauge parameter functions can also depend on the transverse 
coordinates $x^\mu$. In this case, the analogous gauge deformations 
are eaten by the higher Kaluza-Klein gauge fields.

However, this is not the end of the story; the gauge variation of the 
Green-Schwarz counter term (with exact terms neglected) 
is given by \cite{GSW}
\beqa
\delta%
(-dB \wedge X_7)&=&
-\delta%
(dB) \wedge X_7
-dB \wedge \delta%
(X_7) \n
&=&
-d(\omega_{2Y}^1-\omega_{2L}^1) \wedge X_7
-dB \wedge d X_6^1.
\label{delta(GS)}
\eeqa
The first term is a well-known contribution to complete 
the Green-Schwarz mechanism for anomaly cancellation 
in heterotic string theory, while the second term usually vanishes 
by partial integration. However, if there is a magnetic source of the 
$B$ field, the latter also gives a nonzero result since 
\beqa
d^2 B\propto \delta^4(x).
\label{d^2B=delta}
\eeqa
Therefore, unlike the unitary gauge in the usual Higgs mechanism, 
gauge variations of the heterotic 5-brane configuration are not 
completely absorbed into the massive gauge fields.

The clearest evidence for the existence of chiral fermions on the 
brane is provided by the anomaly cancellation argument \cite{anomaly_cancellation,KM2}.
If there are 120 NG bosons on the brane, their quasi NG fermions 
produce gauge and gravitational anomalies. It has been shown in \cite{KM2}
that the sum of these anomalies and the anomaly inflow 
(cf. (\ref{delta(GS)}),(\ref{d^2B=delta})) turns out to be in a factorized form 
just like the ordinary ten-dimensional bulk anomalies, 
and can be similarly cancelled by a Green-Schwarz counter term 
on the brane. If it were not for the 115 zero modes from the spontaneously
broken gauge rotations, the sum of anomalies would fail to factorize, 
and the successful cancellation argument would be invalidated.

To recap, an NS5-brane in heterotic string theory has a gauge instanton 
in the transverse space and spontaneously chooses a particular 
embedding of the spin connection into the gauge connection. 
The associated NG modes are localized on the brane 
together with their superpartners, 
whose anomalies consistently match with the anomaly inflow from the bulk.
This opens up a possibility 
of realizing the nonlinear sigma models on the $E_8$ cosets discussed 
in section \ref{section2}.

\subsection{Intersecting 5-branes and  $E_{8}/(E_6\times SU(2)\times U(1))$
sigma model}\label{Intersecting}
\begin{figure}%
\includegraphics[height=0.3\textheight]{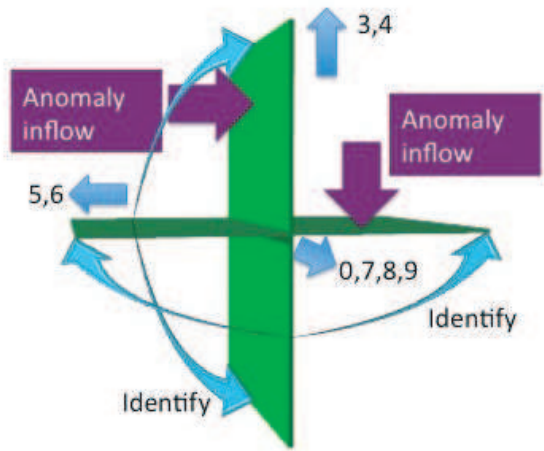}
\caption{\label{Fig:Intersecting} Two stacks of intersecting NS5-branes.
The intersection is four-dimensional. Relatively transverse directions 
are compactified on a torus by periodic identifications. 
The two overall transverse directions ($x^1$ and $x^2$) are not depicted here.}
\end{figure}

In the previous section, we 
have seen that 
a stack of parallel NS5-branes spontaneously breaks the translation, 
scale and gauge rotation symmetries to yield localized modes on 
the magnetic-source locus of the $B$ field.
In order to have, on the other hand, a four-dimensional theory with $N=1$ 
supersymmetry, we consider \cite{KM, KM2,Yata} two stacks of 
intersecting NS5-branes. 
In four dimensions there is no pure gravitational anomaly, 
nor is there gauge anomaly for $E_6$.
However, each of 5-branes should support chiral fermions
so that  
their anomalies cancel against the anomaly inflow from the bulk.
The two 5-branes have four common space-time directions, one is 
extended in $(x^0,x^5,x^6,x^7,x^8,x^9)$ directions and the other 
in $(x^0,x^3,x^4,x^7,x^8,x^9)$ directions.
Thus we compactify the relatively transverse directions $(x^3,x^4,x^5,x^6)$  
on a 4-torus. The configuration that we consider is also smeared 
in one of the overall transverse directions $x^2$, which 
is further compactified on $S^1$. In this way we end up with  
a system of four-dimensional space-time ($x^0,x^7,x^8,x^9$)
with a single extra dimension $x^1$ (FIG.\ref{Fig:Intersecting}; see also FIG\ref{figS1}).

Supergravity solutions for two intersecting NS5-branes  
 localized in all but one of the transverse 
directions are known \cite{McOristRoyston1101.3552}.
They were obtained in type II theories and hence are gauge neutral, but may also 
be promoted to leading-order heterotic supergravity solutions by 
the standard embedding. However, later we are going to analyze the gaugino 
Dirac equation on the intersecting NS5-brane background, and  
this solution is too complicated for our purposes.
Instead, we consider the well-known smeared solution in all except one 
transverse direction \cite{smearedsolution}, 
which is very simple to deal with. 
One of its virtue is that, unlike the original parallel 5-brane solution (\ref{heterotic5}) 
which develops a throat geometry where the theory is strongly coupled, the 
domain-wall type smeared solution has a finite string coupling even near 
the brane locus.  
This  
solution is also interesting since, 
as we mentioned above, it can be reduced 
to a Randall-Sundrum-like \cite{RS}\footnote{However, there is no cosmological 
constant in our setup, either in the bulk or on the branes. Also, warping is 
much smaller. }
extra dimension 
model with a GUT gauge group, a setup similar 
to that postulated in the orbifold GUT or the 
grand gauge-Higgs unification 
scenario \cite{orbifoldGUT,grandGHU}.

In the string frame, the (gauge) neutral solution is given by
\begin{eqnarray}
ds^2&=&\eta_{ij}dx^i dx^j
+h(x)^2\delta_{\mu\nu}dx^\mu dx^\nu
+h(x)\delta_{\mu'\nu'}dx^{\mu'} dx^{\nu'}
\label{intersecting5}
 \\
e^{2\phi}&=&h(x)^2,\nonumber\\
H_{234}&=&H_{256}~=~\frac{h'(x)}2
\nonumber
\eeqa
with
\beqa
h(x)&=&g_0\left(1-\frac{|x|}L\right),
\eeqa
where $i,j$ run over $\{0,7,8,9\}$, $\mu',\nu'$ do over $\{3,4,5,6\}$
and $\mu,\nu$ take values $\{1,2\}$. 
All the fields depend only on the $x^1$ coordinate, and we write $x\equiv x^1$.

The indices of $H$ are the curved ones. Other components of $H$ are 
zero. $g_0$ and $L$ are real, positive constants; $g_0$ is the value of 
the string coupling at $x=0$, whereas $L$ is the scale of the transverse 
dimension.

These set of configurations describe two intersecting NS5-branes, 
stretched along the $x^0$-, $x^5$-, $x^6$-, $x^7$-, $x^8$- and  $x^9$-axes 
and the $x^0$-, $x^3$-, $x^4$-, $x^7$-, $x^8$- and  $x^9$-axes, 
respectively.
They satisfy the equations of motion and 
preserve 1/4 of supersymmetry if 
\beqa
|x|&<&L,
\eeqa
so that the function $h(x)$ is positive;  
the solution is well-defined only in a finite interval 
with a finite proper distance.
In order to avoid the occurrence of a negative $h(x)$ region, 
we consider \cite{KM,KM2,Yata} a periodic array of (\ref{intersecting5}) by taking
\beqa
h(x)&=&g_0\left(1-\frac{| x- 2n L |}L \right)~~~\mbox{if}~~~
(2n - 1)L < x\leq (2n + 1)L  
~~~\mbox{for some}~n\in {\bf Z}
\label{periodic_h(x)}
\eeqa
(FIG.\ref{periodic_h}), and periodically identify the $x(=x^1)$ space 
by a relation
$x\sim  x+2L$. Thus the $x^1$ direction is compactified 
on a circle. (Later we argue that this circle should be ${\bf Z}_2$ orbifolded 
further.)

\begin{figure}%
\includegraphics[height=0.24\textheight]{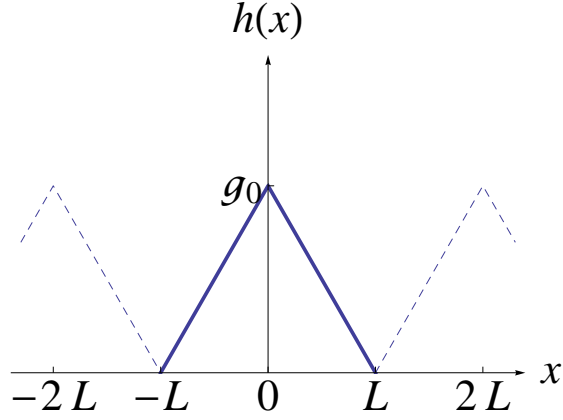}
\caption{\label{periodic_h} The periodic function $h(x)$.
}
\end{figure}

The function $h(x)$ has two singularities, at which magnetically 
charged objects with opposite $H$ charges are located. Note that 
while the one at $x=0$ is an ordinary intersecting 5-branes with 
positive tension, the other at $x=\pm L$ has negative 
tension.  Later, in section \ref{orientifold}, we will discuss how such  
an orientifold-like object can be understood in heterotic string 
theory.

The solution (\ref{intersecting5}) preserve 1/4 of supersymmetry. 
One of the Killing spinor equations is the gravitino variation equation:
\begin{eqnarray}
\delta \psi_M&=&\left(
\partial_M +\frac14 (\omega - H)_{M}{}^{AB}\Gamma_{AB}
\right) \epsilon~=~0.
\label{gravitino_variation}
\end{eqnarray}
The existence of a non-trivial Killing spinor implies that the $SO(6)$
connection $\omega - H$ is actually in $SU(3)$, as is easily verified. 
Also it can be checked that the other combination $\omega + H$ is 
also in a different $SU(3)$ sub-algebra of $SO(6)$. Explicitly,
\beqa
\omega_1+H_1&=&0,\n
\omega_2+H_2&=&
-i\frac{h'}{2h^2}\cdot
\frac{3\lambda_3+\sqrt{3}\lambda_8}2 \otimes \sigma_2,\n
\omega_3+H_3&=&
-i\frac{h'}{2h^2}\cdot
\lambda_2 \otimes {\bf 1}, \label{omega+H}
\\
\omega_4+H_4&=&
-i\frac{h'}{2h^2}\cdot
\lambda_1 \otimes \sigma_2,\n
\omega_5+H_5&=&
-i\frac{h'}{2h^2}\cdot
\lambda_5 \otimes {\bf 1},\n
\omega_6+H_6&=&
-i\frac{h'}{2h^2}\cdot
\lambda_4 \otimes \sigma_2,\nonumber
\eeqa
where the three of the latter two indices of $\omega$ and $H$ are 
regarded as the matrix indices. $\lambda_1,\ldots,\lambda_8$ are 
the Gell-Mann matrices, and $\sigma_1,\sigma_2,\sigma_3$ are the 
Pauli matrices.

Each $2\times 2$ block of the $6\times 6$ matrices (\ref{omega+H}) 
is either ${\bf 1}$ or $i \sigma_2$.
In mapping $SO(6)$ to $SU(3)$, ${\bf 1}$ can be simply dropped, 
while there is a sign ambiguity in reducing $i \sigma_2$ to $\pm i$;
flipping this sign corresponds to going from the ${\bf 3}$ representation 
to the ${\bf \bar 3}$ representation, and vice versa.  
Denoting this sign by $s=\pm$, the $SU(3)$ configuration obtained by 
the standard embedding is given by
\beqa
A_1&=&0,\n
A_2&=&
-i\frac{h'}{2h^2}\cdot
s\frac{3\lambda_3+\sqrt{3}\lambda_8}2 ,\n
A_3&=&
-i\frac{h'}{2h^2}\cdot
\lambda_2 , \label{embeddedA}
\\
A_4&=&
-i\frac{h'}{2h^2}\cdot
s\lambda_1 ,\n
A_5&=&
-i\frac{h'}{2h^2}\cdot
\lambda_5 ,\n
A_6&=&
-i\frac{h'}{2h^2}\cdot
s\lambda_4 .\nonumber
\eeqa

We now consider the moduli. The gauge configuration (\ref{embeddedA}) 
chooses a particular $SU(3)$ sub-algebra of $E_8$. The gauge rotation 
moduli consist of the gauge rotations that change this configuration.
They are those in 
\beqa
 ({\bf 1},{\bf 8}),~~~ ({\bf 27},{\bf 3}),~~~ ({\bf \overline{27}},{\bf \bar 3})
 \label{18-273-273}
\eeqa
as representations of  $E_6 \times SU(3)$.
They are {\em real} coordinates of the coset space $E_8/E_6$, which is 
not K\"{a}hler. However, (\ref{18-273-273}) can be understood as 
arising from a two-step breaking, in which $E_8$ is first broken to 
 $E_6\times U(1) \times U(1)$ (or $E_6\times SU(2) \times U(1)$), 
 giving a K\"{a}hler coset, and 
 then the  $U(1) \times U(1)$ (or $SU(2) \times U(1)$) symmetry
is broken. The second breaking does not give rise to any $E_6$-charged 
fields. Moreover, recently it has been found \cite{KYsupergravity} that 
such extra $U(1)$ degrees of freedom are eliminated from the denominator 
if the sigma models are coupled to supergravity. 
Therefore, we consider the 
nonlinear sigma model on  $E_8/(E_6\times SU(2) \times U(1))$ 
(which is simpler than $E_8/(E_6\times U(1) \times U(1))$; 
the chiral spectrum does not change whatever $Y$-charge is chosen) 
and examine what it implies about the chiralities of massless 
fermions on this background.

$E_8/(E_6\times SU(2) \times U(1))$ is the last example discussed 
in section \ref{section2}, where we have seen that the sigma model 
has three chiral quasi NG fermions transforming nontrivially by 
the $E_6$ rotation. They are:
\beqa
({\bf 27},{\bf 2}) ~~\mbox{and}~~({\bf \overline{27}},{\bf 1}).
\eeqa
Thus the sigma model predicts that there are two chiral and one 
anti-chiral generations transforming as ${\bf 27}$ of $E_6$. In the next 
section, we confirm this prediction by explicitly solving the gaugino 
Dirac equation on the smeared intersecting 5-brane background. 
Essentially, this result was  already obtained and announced in 
\cite{KM,KM2}; in the present paper we will give a more complete 
discussion, including the precise boundary conditions set for the solutions 
and (non)normalizability
\footnote{In fact, it is not obvious whether the existence of zero modes 
of the gaugino equation of motion, to be discussed in the next section, directly 
implies the existence of modes localized on the brane; according to 
the anomaly argument in section \ref{What_symmetry}, 
the latter is supposed to live 
on the $\delta$-function-like B-field source, while the former has a profile extended into 
the transverse dimensions. 
However, since the modes on the brane 
arise as a boundary contribution due to (\ref{d^2B=delta}), 
it is natural to expect that the chirality of the bulk gaugino solution 
and that of the localized mode coincide, which we assume 
in this paper. 
Note, however, that this is the conventional interpretation of the zero mode in the 
literature (e.g. \cite{CHinflow}).}.

\subsection{Explicit computation of fermionic zero modes}
\label{Explicit}
The gaugino equation of motion of heterotic string theory can be 
compactly written in terms of a special combination of the spin and
gauge connections (see e.g. \cite{KYi}):
\beqa
\Gamma^M D_M\left(
\omega - \frac H3, A
\right)\tilde\chi &=&0,
\eeqa
where
\beqa
\chi=e^\phi \tilde \chi 
\eeqa
is the original gaugino in the Lagrangian.

\beqa
\Gamma^i D_i \tilde\chi+\Gamma^\mu D_\mu\left(
\omega - \frac H3, A
\right)\tilde\chi &=&0
\label{10DgauginoEOM}
\eeqa

\beqa
\tilde \chi &=& \tilde \chi_{4D}(x^i) \otimes \tilde \chi_{6D}(x^\mu)
\eeqa

See Appendix \ref{conventions} for conventions for the gamma matrices used in this paper.
In these conventions, a Weyl spinor with $\Gamma_{11}=+1$ is a linear combination of the 
spinors of the forms
\beqa
{\tiny
\left[
\begin{array}{c}
*\\ * \\0\\0
\end{array}
\right] \otimes
\left[
\begin{array}{c}
*\\ * \\ * \\ * \\0\\0\\0\\0
\end{array}
\right]} 
~\mbox{and}~
{\tiny
\left[
\begin{array}{c}
0\\0 \\ * \\ * 
\end{array}
\right] \otimes
\left[
\begin{array}{c}
0\\0\\0\\0\\ * \\ * \\ * \\ * 
\end{array}
\right]
}.
\eeqa

The gaugino is Majorana-Weyl
in ten dimensions. In the present notation a spinor $\psi$ is said Majorana 
iff  
\beqa
B\psi^*&=&\psi,
\eeqa
where ${}^*$ denotes the complex conjugate. Now suppose that 
we found $\chi_{6D}$ of the form 
{\tiny $\left[
\begin{array}{c}
*\\ * \\ * \\ * \\0\\0\\0\\0
\end{array}
\right] $}
satisfying
\beqa
\gamma^\mu D_\mu\left(
\omega - \frac H3, A
\right)\tilde\chi_{6D} &=&0
\label{6Dgauginoeq}
\eeqa
and $\chi_{4D}$ of the form 
{\tiny $\left[
\begin{array}{c}
*\\ * \\0\\0
\end{array}
\right] 
$}  satisfying
\beqa
\gamma^i_{4D} D_i \tilde\chi_{4D}&=&0.
\eeqa
Then we can construct a Majorana spinor 
\beqa
\tilde \chi + B \tilde \chi^*, ~~~ \tilde\chi=\tilde\chi_{4D} \otimes \tilde\chi_{6D}.
\eeqa
which satisfies the equation 
(\ref{10DgauginoEOM}). Thus, to look for gaugino zero modes, it is enough to 
solve the equation (\ref{6Dgauginoeq}) 
for $\tilde\chi_{6D}$ of the form 
\beqa
\tilde\chi_{6D}&=&
\left(
\begin{array}{c}
\tilde\chi_{6D}^+ \\ 0
\end{array}
\right),
\label{tildechi6D}
\eeqa
where $\tilde\chi^+$ is a four-component spinor.  
Note that if we consider $\tilde\chi_{6D}$ transforming as $({\bf 27},{\bf 3})$,
each of the four components of $\tilde\chi_{6D}$ is tensored by 
a triplet vector of $SU(3)$ (and also a 27-plet of $E_6$).

We consider $\tilde\chi_{6D}$ in $({\bf 27},{\bf 3})$ that depends only on the 
$x^1(\equiv x)$ 
coordinate.
Writing 
\beqa
\gamma^1=\left(
\begin{array}{cc}
& -i{\bf 1}_4\\
i{\bf 1}_4 &
\end{array}
\right),~~~
\gamma^{\tilde\alpha}=\left(
\begin{array}{cc}
& \tilde\gamma^{\tilde\alpha}\\
\tilde\gamma^{\tilde\alpha} &
\end{array}
\right)~~~(\tilde\alpha=2,\ldots,6),
\eeqa
the equation (\ref{6Dgauginoeq}) becomes
\beqa
\left(
h^{-1}\left(
\begin{array}{cc}
&-i \partial_x \\
i\partial_x&
\end{array}
\right)
+\frac14\left(
\omega -\frac H 3
\right)_{\tilde\alpha \beta\gamma} \gamma^{\tilde\alpha \beta\gamma}
+
\left(
\begin{array}{cc}
&A_{\tilde\alpha} \tilde\gamma^{\tilde\alpha}\\
A_{\tilde\alpha}\tilde\gamma^{\tilde\alpha} &
\end{array}
\right)
\right)\tilde\chi_{6D}(x)&=&0,
\eeqa
where the second term is also in the block off-diagonal form.
This yields a simple first-order differential equation for $\tilde\chi_{6D}^+$:
\beqa
\left(\frac d{dx}+ \frac{h'(x)}{h(x)} M(s) \right) \tilde\chi_{6D}^+ &=&0,
\eeqa
where the matrix $M(s)$ is a real, constant matrix given by
%
%
\beqa
M(s)&=&
\left(
\begin{array}{cccccccccccc}
 \frac{3}{2} & 0 & -\frac{s}{2} & -\frac{1}{4} & -\frac{s}{2} & \frac{1}{2} & -s & \frac{1}{2} & 0 & -\frac{1}{4} & 0 & 0 \\
 0 & \frac{3}{2} & 0 & -\frac{s}{2} & -\frac{1}{4} & 0 & -\frac{1}{2} & \frac{s}{2} & 0 & 0 & -\frac{1}{4} & 0 \\
 -\frac{s}{2} & 0 & \frac{3}{2} & -\frac{1}{2} & 0 & -\frac{1}{4} & 0 & 0 & \frac{s}{2} & 0 & 0 & -\frac{1}{4} \\
 -\frac{1}{4} & -\frac{s}{2} & -\frac{1}{2} & \frac{3}{2} & 0 & \frac{s}{2} & -\frac{1}{4} & 0 & 0 & -s & \frac{1}{2} & 0 \\
 -\frac{s}{2} & -\frac{1}{4} & 0 & 0 & \frac{3}{2} & 0 & 0 & -\frac{1}{4} & 0 & -\frac{1}{2} & \frac{s}{2} & 0 \\
 \frac{1}{2} & 0 & -\frac{1}{4} & \frac{s}{2} & 0 & \frac{3}{2} & 0 & 0 & -\frac{1}{4} & 0 & 0 & \frac{s}{2} \\
 -s & -\frac{1}{2} & 0 & -\frac{1}{4} & 0 & 0 & \frac{3}{2} & 0 & \frac{s}{2} & -\frac{1}{4} & \frac{s}{2} & -\frac{1}{2} \\
 \frac{1}{2} & \frac{s}{2} & 0 & 0 & -\frac{1}{4} & 0 & 0 & \frac{3}{2} & 0 & \frac{s}{2} & -\frac{1}{4} & 0 \\
 0 & 0 & \frac{s}{2} & 0 & 0 & -\frac{1}{4} & \frac{s}{2} & 0 & \frac{3}{2} & \frac{1}{2} & 0 & -\frac{1}{4} \\
 -\frac{1}{4} & 0 & 0 & -s & -\frac{1}{2} & 0 & -\frac{1}{4} & \frac{s}{2} & \frac{1}{2} & \frac{3}{2} & 0 & -\frac{s}{2} \\
 0 & -\frac{1}{4} & 0 & \frac{1}{2} & \frac{s}{2} & 0 & \frac{s}{2} & -\frac{1}{4} & 0 & 0 & \frac{3}{2} & 0 \\
 0 & 0 & -\frac{1}{4} & 0 & 0 & \frac{s}{2} & -\frac{1}{2} & 0 & -\frac{1}{4} & -\frac{s}{2} & 0 & \frac{3}{2}
\end{array}
\right),
\eeqa
where $s=\pm 1$ distinguishes which of ${\bf 3}$ or ${\bf \bar 3}$ the solution belongs to.
This matrix can readily be diagonalized, and the differential equation can be easily 
solved \cite{KM,KM2}. The eigenvalues are
\beqa
-1  ,~1,~  1,~  1,~  1,~ \frac3 2,~\frac3 2,~ \frac3 2,~\frac3 2,~ 2   ,~\frac7 2 ,~\frac7 2
\label{s=+1}
\eeqa
if $s=+1$, and
\beqa
-\frac12,-\frac12,~  1,~ \frac3 2,~\frac3 2,~ \frac3 2,~\frac3 2,~ 2, ~ 2,~ 2,~ 2  ,~4
\label{s=-1}
\eeqa
if $s=-1$.
An intriguing feature of these two sets of eigenvalues is that 
they are mapped with each other by the reflection about $\frac32$:
$\lambda\mapsto 3-\lambda$.

\begin{figure}[b]%
\includegraphics[height=0.3\textheight]{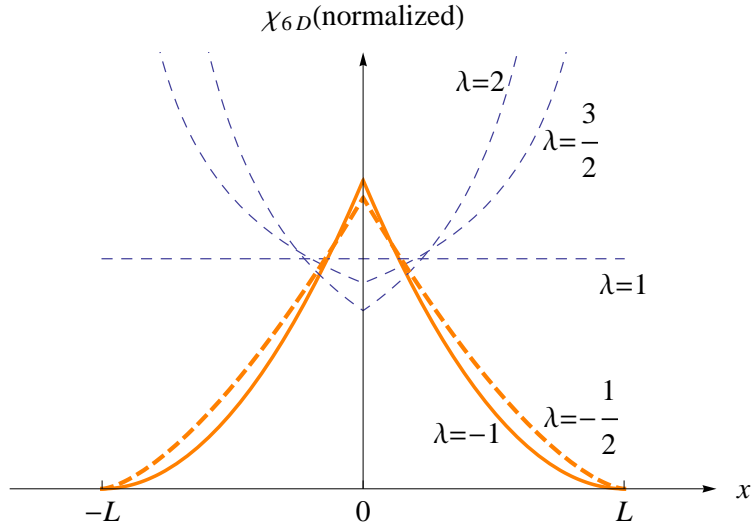}
\caption{\label{zeromode_profiles} The normalized zero mode profiles.
}
\end{figure}

Suppose that $\lambda$ is one of the eigenvalues of $M$ above, and let $\eta_\lambda$ 
be a corresponding (constant) eigenvector:
\beqa
M\eta_\lambda&=&\lambda\eta_\lambda.
\eeqa
Then, writing 
\beqa
\tilde\chi_{6D}^+&=&f_\lambda(x)\eta_\lambda
\eeqa
with some scale factor function $f_\lambda(x)$, $f_\lambda(x)$ is trivially solved 
to be
\beqa
f_\lambda(x)&=&\mbox{constant}\times~h(x)^{-\lambda}
\eeqa
so that
\beqa
\tilde\chi_{6D}^+&=&\mbox{constant}\times~h(x)^{-\lambda}\eta_\lambda
\eeqa
or
\beqa
\chi_{6D}&=&\mbox{constant}\times~h(x)^{-\lambda+1}
\left(\begin{array}{c}
\eta_\lambda\\
0
\end{array}
\right).
\eeqa

In any zero-mode analysis, it is crucial to set appropriate boundary conditions for the 
solutions. In the present case, we are interested in the modes localized near $x=0$,
the locus of intersecting 5-branes with positive tension, and not near $x=\pm L$ 
where the negative tension ones reside. Thus we impose that the solution should vanish 
at $x=\pm L$.
This boundary condition requires that the eigenvalues of $M$ must be  
less than 1,
which are $-1$ for $s=+1$ and two $-\frac12$'s for $s=-1$ (FIG.\ref{zeromode_profiles}). 
Therefore, 
(calling the former ``${\bar{\bf 3}}$" and the latter ``${\bf 3}$" of $SU(3)$)
we find two chiral and one anti-chiral zero modes that satisfy the 
boundary condition, 
being 
in agreement with the prediction made by the 
$E_8/(E_6\times SU(2) \times U(1))$
or
$E_8/(E_6\times U(1) \times U(1))$
supersymmetric sigma model.

\subsection{Zero modes near the negative tension brane and (non)normalizability}
Other eigenvalues $\lambda$, the ones except $-1$ in (\ref{s=+1}) and two $-\frac12$'s in (\ref{s=-1}), 
are all larger then or equal to $1$. The overall function in $\chi_{6D}$ blows up at 
$x=\pm L$
if $\lambda>1$, or are constant if $\lambda=1$.  

On the background with the string frame metric (\ref{intersecting5}), the volume form $d\mu$
of the transverse dimensions is given by
\beqa
d\mu&=& dx^1 dx^2 \cdots dx^6 (h(x^1))^4.
\eeqa
If the delocalized dimensions $x^2,\ldots,x^6$ are compactified on $T^5$ with 
a common radius $R$, then the measure becomes
\beqa
d\mu&=&dx\cdot (2\pi R)^5 h(x)^4,
\label{dmu}
\eeqa
where $x^1\equiv x$.
Therefore, in order for $\chi_{6D}$ to be normalizable:
\beqa
\int d\mu \chi_{6D}^\dag \chi_{6D}&<&\infty,
\eeqa
the eigenvalue $\lambda$ must satisfy
\beqa
\lambda&<&\frac72.
\eeqa
(If $\lambda=\frac72$ the norm diverges logarithmically.)
The modes that do not satisfy this condition is the two modes 
with $\lambda=\frac72$ for $s=+1$, and the one 
with $\lambda=4$ for $s=-1$. Thus there are three non-normalizable modes,
two are anti-chiral and one is chiral. This leaves an equal number of chiral 
and anti-chiral 
zero modes that are {\em not} localized near $x=0$.

In view of this, one 
might expect that they would be grouped into pairs and become massive. 
However, these remaining set of eigenvalues in (\ref{s=+1}) and (\ref{s=-1}) 
are not completely equal. For instance, (\ref{s=+1}) has four 1's, 
whereas  (\ref{s=-1}) has only a single such mode. Since 
different eigenvalues correspond to different profiles, 
this fact makes   
it questionable whether these modes are all grouped into massive modes.
If this $(2-1)$-generation toy model is generalized to be more realistic 
for phenomenological applications, then 
the effect of these modes  
would also need to be seriously considered.  
 
\subsection{An orientifold substitute in heterotic string theory}
\label{orientifold}

In section \ref{Intersecting}, we considered a periodic harmonic 
function in the 5-brane solution so that one of the overall transverse 
dimensions be compactified.
The periodic harmonic function necessarily led to the introduction of 
negative tension branes.
In general, 
it is a well-known fact that for warped compactification there must be 
some branes with negative tension \cite{GKP}, so
what we have encountered above may be seen to be 
consistent with the no-go theorem.

In type II theories 
an orientifold can be identified as such a negative tension object. 
However, in heterotic string theory, 
the ordinary microscopic definition of orientifolds does not make 
sense because the heterotic string world sheet is asymmetric.  
How can we understand such a negative tension object in heterotic string theory?

This problem was considered in \cite{T-NUTcrystal} (see also \cite{AHTdual} for earlier 
discussions), where it was argued that a negative tension brane in 
heterotic string theory could be understood as a T-dual of the Atiyah-Hitchin \cite{AH}
manifold. The argument is based on the fact that the metric of the 
Atiyah-Hitchin space asymptotically approaches that of the Taub-NUT space but 
with {\em negative} NUT charge at large distances.  
More precisely, the Atiyah-Hitchin metric can be written,  by using elliptic 
theta functions, in the form 
\beqa
&&ds^2=a^2 b^2 c^2 dt^2 + a^2 \sigma_1^2+ b^2 \sigma_2^2+ c^2 \sigma_3^2,\n
&&a^2=\frac{w_2 w_3}{w_1},~~~b^2=\frac{w_3 w_1}{w_2},~~~c^2=\frac{w_1 w_2}{w_3},\\
&&w_j=2\frac d{dt}\log \vartheta_{j+1}(0,2 \pi i m^2 t)~~~(j=1,2,3)
\eeqa 
with $\sigma_1,\sigma_2$ and $\sigma_3$ being the $SU(2)$ Maurer-Cartan 1-forms.
If one ignores the rapidly vanishing terms containing $e^{-\frac 1{2 m^2 t} }$ 
in the theta functions, 
one can obtain the asymptotic behavior of the metric: 
\beqa
ds^2&\stackrel{t\rightarrow 0}\sim&
\frac{1-4 m^2 t}{4 m^2 t^4} dt^2
+\frac{4 m^2}{1 - 4 m^2 t} \sigma_3^2 
+ \frac{1- 4 m^2 t}{4 m^2 t^2} (\sigma_1^2 + \sigma_2^2)\n
&=&\left(
1-\frac{2m}r
\right)
(dr^2 + r^2(\sigma_1^2 + \sigma_2^2))
+4m^2
\left(
1-\frac{2m}r
\right)^{-1}
\sigma_3^2,
\label{negativeT-NUT}
\eeqa
where $r=\frac 1{2mt}$ $(t>0,m>0)$. 
This is the negative-charge Taub-NUT. 
While the Atiyah-Hitchin space is a smooth manifold,
a singularity at $r=2m$ has 
arisen in the 
asymptotic form (\ref{negativeT-NUT}) 
because the infinitely many exponential terms has been 
discarded. From the probe gauge theory point of view, 
these terms are regarded as the instanton corrections \cite{T-NUTcrystal}.

The Atiyah-Hitchin metric satisfies Einstein's equation, and so does 
the asymptotic form (\ref{negativeT-NUT}). One can embed  (\ref{negativeT-NUT}) 
in ten dimensions as a part of a string frame metric. Then by taking T-duality 
along the Hopf fiber direction of the Taub-NUT, we obtain 
\beqa
ds_{\mbox{\scriptsize T-dual}}^2&=&
\eta_{ij} dx^i dx^j 
+
V(r) \delta_{\mu\nu} dx^\mu dx^\nu,
\n
e^{2\Phi^{\mbox{\scriptsize T-dual}}}&=&V(r),\label{smearedNS5}
\\
H^{\mbox{\scriptsize T-dual}}_{\mu\nu 9}&=&\epsilon^\rho_{~\mu\nu 9 }\partial_\rho V(r)
\nonumber
\eeqa
where
\beqa
V(r)&=&1 - \frac{2m}r 
\label{negativeV(r)}
\eeqa
with $r$ being the {\em three}-dimensional radial coordinate here.
Note that the ordinary smeared 5-brane has $V(r)=1+\frac{2m}r$.
By further delocalizations except one direction the three-dimensional 
harmonic function $V(r)$ is replaced with
\beqa
1-\frac{2m}r&\rightarrow&1-2m|x|.
\label{1dharmonic}
\eeqa
Since the smeared intersecting solution (\ref{intersecting5}) 
can be viewed as a superposition of two smeared solutions of the 
form (\ref{smearedNS5}), the negative tension objects located at 
$x=\pm L(+2nL)$ (FIG.\ref{periodic_h}) could be thought of as a 
superposition of two T-duals of the Atiyah-Hitchin manifold \cite{T-NUTcrystal}.

While negative tension branes in heterotic string theory sounds bizarre, 
heterotic strings propagating on the celebrated Atiyah-Hitchin 
hyper-K\"{a}hler manifold will be no problem.  Therefore, if  
a smooth T-dual (or mirror) of the Atiyah-Hitchin space exists, the 
negative tension branes will be asymptotic approximation of it at large 
distances. Also, since a T-dual of two intersecting 5-branes (with positive tensions)
is a conifold \cite{conifold,McOristRoyston1101.3552}, it would be interesting to explore 
what corresponds to the Atiyah-Hitchin manifold in six dimensions.

\begin{figure}[b]%
\includegraphics[height=0.2\textheight]{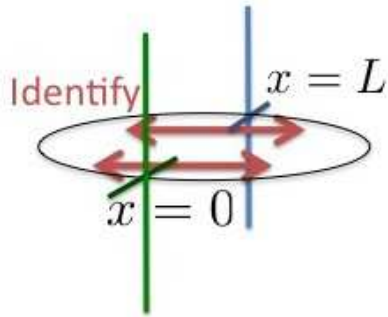}
\caption{\label{figS1} The ${\bf Z}_2$ identification due to the ``bolt".
}
\end{figure}

An important aspect of this identification is that the transverse space 
is {\em forced} to be ${\bf Z}_2$-orbifolded about the loci of the 
negative tension branes, similarly to the ordinary orientifolds 
in type II theories. This is because the Atiyah-Hitchin manifold 
has a so-called ``bolt" \cite{EGH} at the center \cite{GibbonsManton}. 
Therefore, although at the beginning we started from the $x(=x^1)$ space 
compactified on a circle by taking a periodic array (\ref{periodic_h(x)}), 
the circle is necessarily subject to the ${\bf Z}_2$ identifications 
\beqa
x  &\sim& 2(2n +1)L -x~~~(n\in{\bf Z}),
\eeqa
so that the circle is orbifolded into an interval, which is obligatory 
for our setup (FIG.\ref{figS1}). This is in contrast to what is usually done in 
the gauge Higgs unification \cite{GHU} 
or orbifold GUT theories \cite{orbifoldGUT,grandGHU}, where the ${\bf Z}_2$
projection is imposed {\em by hand}. 
Thus our intersecting 5-brane scenario may %
provide a natural origin of  
the ${\bf Z}_2$ orbifoldization in the extra dimensions, which has 
been simply postulated in those bottom-up approaches.

\section{Intersecting 5-branes with Wilson lines}\label{Wilson_line}

In the previous section, we have seen that the intersecting 5-branes 
with the $SU(3)$ gauge field obtained by the standard embedding 
support two chiral and one anti-chiral fermionic zero modes, which 
agrees with the prediction of the 
supersymmetric nonlinear 
sigma model on $E_8/(E_6\times U(1) \times U(1))$ 
(or $E_8/(E_6\times SU(2) \times U(1))$).
In order to realize more realistic models with an unbroken 
$SO(10)$ or $SU(5)$ gauge group, we need to 
consider the gauge field configuration 
larger than $SU(3)$.
This amounts to considering a non-standard embedding 
version of the symmetric 5-brane
and their intersection,
which is not known at present.
Therefore, 
as a preliminary investigation into aspects of the 
non-standardly embedded branes, 
we consider in this section a constant Wilson line in the 
transverse space and  see what happens to the profiles 
of the fermion zero modes. 
A discussion on more general gauge configurations  
will be given in the Summary and Discussion section.

Let us consider a constant 
$U(1)$ Wilson line proportional to  
$h_\perp$ $(= \frac13\left(
-(E^1_{~1}+\cdots E^5_{~5})+2 E^6_{~6}+(E^7_{~7}+E^8_{~8}+E^9_{~9})
\right)$
(\ref{h_perp})  in the $x^2$ component of $A_M$ 
in addition to the $SU(3)$ gauge field (\ref{embeddedA}), where 
we take the generators for the latter to be $X_{ab}$, $Y_{ab}$ and $h_{\bar a}$ 
(\ref{SU(3)subalgebra}). The Wilson line breaks the $E_6$ 
gauge symmetry to $SO(10)\times U(1)$. 
Of course, being constant, it does not affect the Bianchi identity (\ref{Bianchi_H}).
According to the decomposition 
of the $E_6$ adjoint 
\beqa
{\bf 78}&=& {\bf 45}_0\oplus {\bf 16}_1 \oplus \overline{\bf 16}_{-1}\oplus {\bf 1}_0
\eeqa
as $SO(10)\times U(1)_\perp$ representations, 
the $E_6$ adjoint gaugino gives rise to 
a ${\bf 16}$ and a $\overline{\bf 16}$ $SO(10)$ gauginos.
They have opposite $U(1)_\perp$ charges.
We write 
\beqa
A_\mu^{U(1)}&=&i w |x|' \delta_\mu^2
\label{AmuU(1)}
\eeqa
with a real constant $w$,
where $|x|' = \frac d{dx}|x| = \pm 1$.
The equation of motion for the ${\bf 16}$ gaugino 
\beqa
\gamma^\mu \left(
\partial_\mu + \frac14\left(
\omega -\frac H3
\right)_{\mu\alpha\beta}
\gamma^{\alpha\beta}
+A^{U(1)}_\mu
\right)
\tilde{\chi}_{6D}&=&0.
\eeqa
is reduced to
\beqa
h^{-1} \left(
\begin{array}{cc}
&-i \partial_{x^1}\\
i \partial_{x^1}&
\end{array}
\right)
+\frac14\left(
\omega+\frac H3
\right)_{\tilde{\alpha}\beta\gamma}\gamma^{\tilde{\alpha}\beta\gamma}
+ i w |x|' h^{-1}\left(
\begin{array}{cc}
&\tilde{\gamma}^2
\\
\tilde{\gamma}^2&
\end{array}
\right)&=&0.
\eeqa
The one for $\overline{\bf 16}$ can be obtained by flipping  
the sign of $w$  (\ref{AmuU(1)}).
Assuming again that $\tilde{\chi}_{6D}$ has only the upper component 
$\tilde{\chi}_{6D}^+$
like  (\ref{tildechi6D}),
the problem amounts to solving 
\beqa
\left(\frac d{dx}+ M^{U(1)} \right) \tilde\chi_{6D}^+ &=&0,
\label{U(1)diffeq}
\eeqa
where $M^{U(1)}$ is this time a $4\times 4$ matrix
\beqa
M^{U(1)}&=&
\left(
\begin{array}{cccc}
 \frac{3 \alpha }{2} & -\frac{\alpha }{4} & c & -\frac{\alpha }{4} \\
 -\frac{\alpha }{4} & \frac{3 \alpha }{2} & -\frac{\alpha }{4} & c \\
 c & -\frac{\alpha }{4} & \frac{3 \alpha }{2} & -\frac{\alpha }{4} \\
 -\frac{\alpha }{4} & c & -\frac{\alpha }{4} & \frac{3 \alpha }{2}
\end{array}
\right)
\eeqa
with $\alpha=\frac{h'}h$ and $c=w |x|'$. This matrix can be 
diagonalized by a {\em constant} matrix $V$:
\beqa
V^{-1} M^{U(1)} V&=& \mbox{diag}\left\{
 \alpha + c, 2\alpha +c, \frac{3\alpha}2 -c, \frac{3\alpha}2 -c
\right\},\\ 
V&=&
\left(
\begin{array}{cccc}
  1&  -1 & 0 & -1\\
  1&   1&   -1&0\\
  1&   -1&   0&1\\
  1&    1& 1&0
\end{array}
\right).
\eeqa
\begin{figure}[ht]%
\includegraphics[height=0.2\textheight]{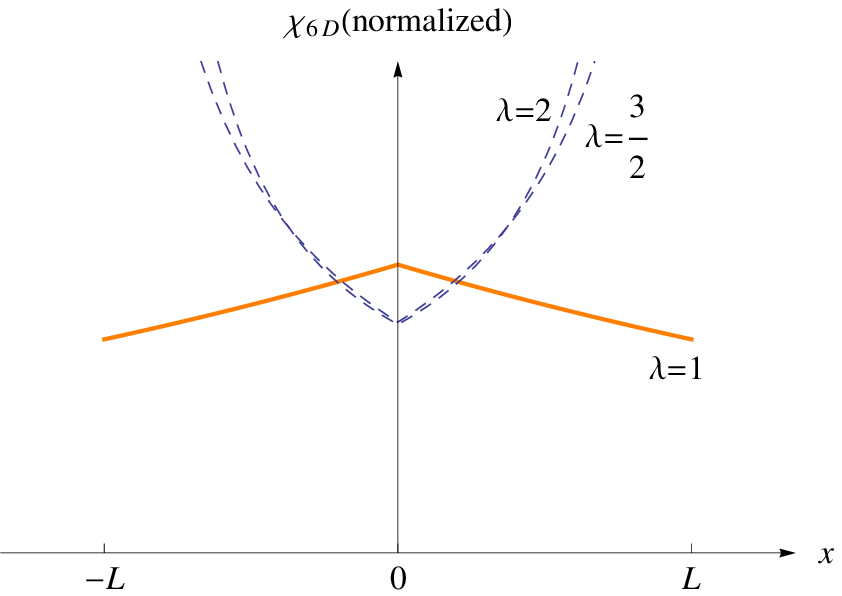}
\includegraphics[height=0.2\textheight]{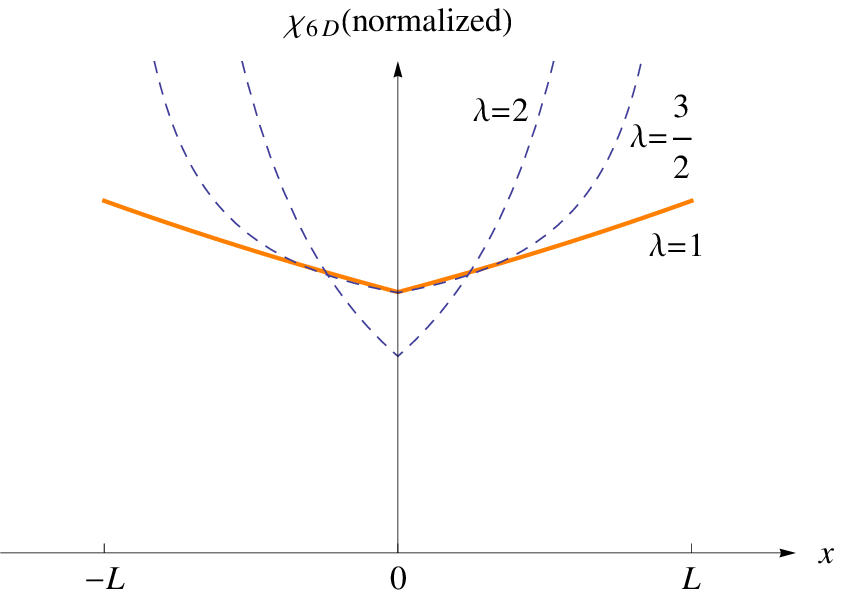}\\
(a-1)\hspace{15.5em}(a-2)~~

\vspace{2em}
\includegraphics[height=0.2\textheight]{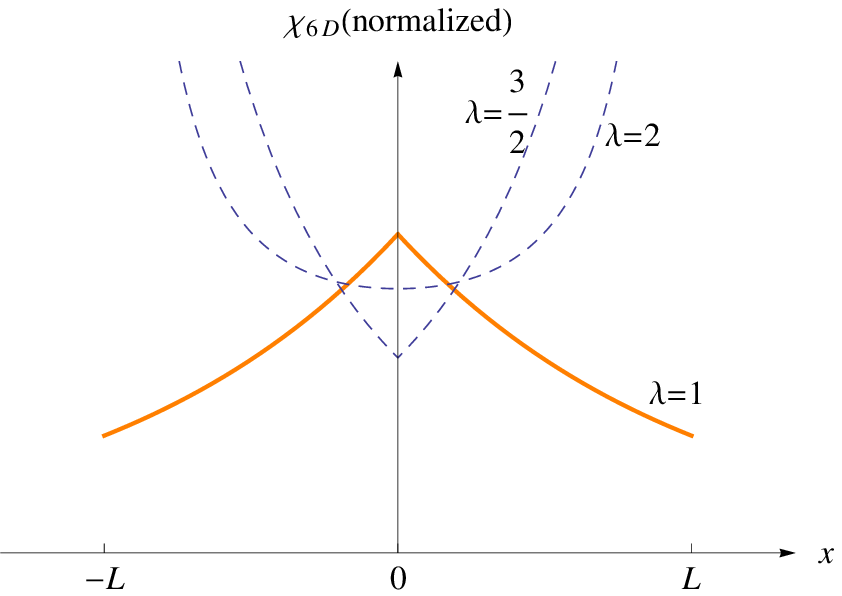}
\includegraphics[height=0.2\textheight]{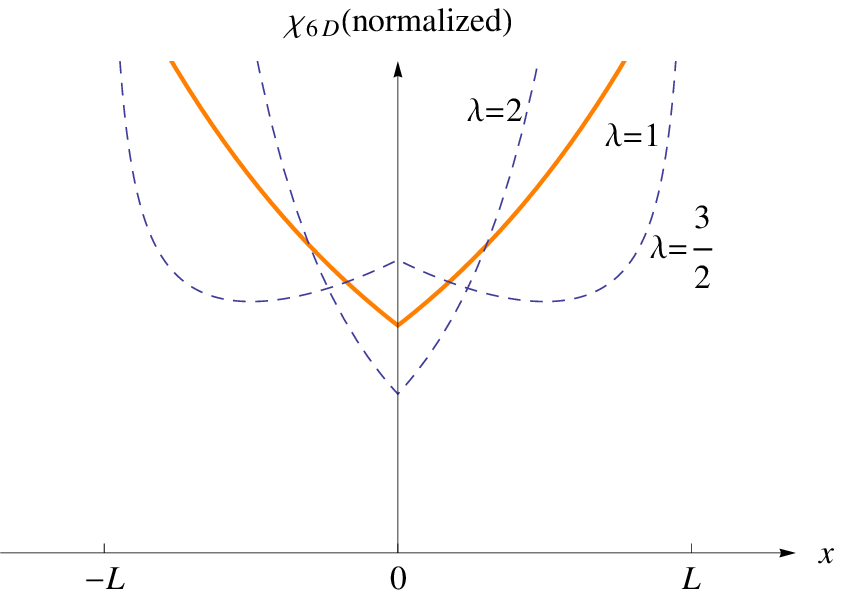}\\
(b-1)\hspace{15.5em}(b-2)~~

\caption{\label{zeromodeU(1)_profiles} The normalized $U(1)_\perp$-charged zero mode profiles.
The cases for $w=0.3$,$-0.3$,$+1$ and $-1$ are shown in (a-1),(a-2),(b-1) and (b-2), respectively.
}
\end{figure}
Note that this simplification does not occur when the Wilson line 
is taken in any of the $x^3,\ldots,x^6$ directions. 
This is one of the reasons why we have chosen $x^2$ as the 
direction of the Wilson line. (Also if $x^1$ is chosen, the 
(correspondingly modified) matrix $M^{U(1)}$ is again diagonalized by a 
constant matrix.)
In the present case, 
(\ref{U(1)diffeq}) is thus easily solved: Let $\lambda=1,2,\frac32$ be 
the coefficient of $\alpha$ of each eigenvalue, the unnormalized 
wave functions are given by
\beqa
\chi^{\lambda,{\rm unnorm.}}_{6D}
&=&h(x)^{-\lambda+1} e^{-(-1)^{2\lambda} w |x|} \eta_\lambda,
\eeqa
where $\eta_\lambda$ is the corresponding column vector of $V$.
Taking  the measure (\ref{dmu}) into account, the profiles (scale factors) 
of the normalized zero mode 
wave functions are shown in FIG.\ref{zeromodeU(1)_profiles}
for $w=\pm0.3$ and $w=\pm 1$.

Let us now compare this result with the coset sigma models. In the present case, 
the relevant coset is $E_8/(SO(10)\times H\times U(1))$ with $H$ being $SU(3)$ or 
$SU(2)\times U(1)$ or $U(1)\times U(1)$.
It is not obvious how the transverse gauge configuration is related to 
the $U(1)$ $Y$-charge of the sigma model. 
In any case, 
however, one can have at most three chiral generations
with an anti-chiral one whatever option of $Y$-charge is made, 
as we wrote in section \ref{E8/(SO(10)xSU(3)xU(1))}. 
Therefore, since we already have two chiral and one anti-chiral $SU(3)$-charged 
zero modes, another ($U(1)_\perp$-charged) chiral zero mode should appear if the 
correspondence to the sigma model holds true. 

In FIG.\ref{zeromodeU(1)_profiles} we can see that the $\lambda=1$ curve changes its 
shape considerably depending on the sign of the Wilson line, while the curves for 
$\lambda=2$ and $\lambda=\frac32$ do not. In particular, only when $\lambda=1$ 
and $w>0$ the profile has the maximum at $x=0$ and monotonically decreases as 
$x\rightarrow \pm L$
\footnote{The $SU(3)$ charged modes found in section \ref{Explicit} are 
also charged with respect to $U(1)_\perp$, but their chiralities  
will not change as far as $g_0$ is small and hence $A_\mu$'s (\ref{embeddedA}) 
are large.}.
However, it is hard to regard 
this mode 
as the expected $U(1)_\perp$-charged zero mode 
because the one with opposite chirality ($\lambda=1$, $w<0$) localizes 
near the other brane at $x=\pm L$. 
Thus we conclude that the inclusion of a constant Wilson line only has the effect of
splitting the chiral and anti-chiral components apart and localizing them 
to regions near the different branes, 
creating no net chiral fermions in the four-dimensional sense.
This means that a constant Wilson line is not enough to realize the $SO(10)$ 
theory in this setup but we need to do something else.

\section{Summary and Discussion}
\label{summary}
Motivated by the fact that 
some exceptional supersymmetric coset sigma models
yield a set of matter fields close to what we really observe in nature, 
we have proposed a possible scenario of realizing 
such a sigma model in heterotic string theory by using 
NS5-branes.
We have considered a system of two stacks of intersecting NS5-branes 
and identified as the one realizing the $E8 / (E6\times SU(2)\times U(1))$ 
(or $E8 / (E6\times U(1)\times U(1))$) sigma model. 
We have examined the validity of the identification by explicitly 
computing the chiral zero modes on the smeared background with 
all the transverse dimensions being compactified except one extra 
dimension. We have found two chiral and one anti-chiral zero modes 
which have their maximum at the location of the positive tension 
brane and vanish at the negative tension brane.  This result is 
consistent with the sigma model prediction, confirming the validity of 
the identification.  

We have also performed a similar computation with including a 
constant Wilson line in one of the transverse dimensions, and 
compared with the spectrum with the $E_8/(SO(10)\times H\times U(1))$ 
sigma model. 
In this case as well, we have indeed found 
a pair of chiral and anti-chiral modes which are asymmetrically 
localized on different branes, but 
it does not seem likely that they can be used as the realization 
of a new chiral generation.
In order to achieve a more realistic coset  like   
$E8/(SO(10)\times SU(3)\times  U(1))$
or 
$E8/(SO(5)\times SU(3)\times  U(1) \times U(1))$, 
a non-standard embedding analogue of the 
NS5-brane will be needed.

In the presence of $H$ fluxes, 
the Killing spinor equation (\ref{gravitino_variation}) 
ensures that $\omega - H$ is in $SU(3)$, but says nothing about 
$\omega + H$. If $dH=0$, it is the latter that is set equal to the gauge 
connection in the ``standard" embedding, and not the former. 
Therefore, the gauge configuration determined 
by setting equal to $\omega + H$ 
{\em need not} be in $SU(3)$ in the presence 
of fluxes, but  
is generically in $SO(6)(=SU(4))$. This is in contrast to the case without 
$H$ fluxes, where there is no distinction between these two ($\omega \pm H$)
connections. 
Nevertheless, for the intersecting 5-brane solution (\ref{intersecting5}) not only $\omega-H$ 
but also $\omega+H$ belongs to $SU(3)$.

For smooth Calabi-Yau manifolds without $H$ flux, it has been known 
for a long time that the Donaldson-Uhlenbeck-Yau theorem ensures 
the existence of an $SU(4)$ gauge field that preserves SUSY 
for every complex structure of a stable holomorphic vector bundle 
\cite{GSW,
WittenB268(1986)79}, and it was suggested that 
such a connection could be obtained by first considering a direct 
sum of the $SU(3)$ bundle and a trivial line bundle, and then 
deforming the complex structure \cite{WittenB268(1986)79}.
It would be natural to ask whether an analogous theorem holds 
in the presence of $H$ fluxes. 
If only the existence of such a brane configuration could be 
ensured, one could in principle use the nonlinear sigma model 
to describe low-energy physics. %

We have argued that the negative tension brane needed 
in the global model could be understood 
as a superposition of T-duals of the Atiyah-Hitchin manifold.
The setup has been reduced to a system similar to that 
postulated in the gauge-Higgs unification scenario,
where the $Z_2$ orbifold structure is naturally inherited from the bolt 
singularity in the Atiyah-Hitchin 
manifold.
 It would be interesting to study the Hosotani mechanism 
in this setup.
Also, a microscopic understanding of the system of 
intersecting NS5-branes is desirable. Since it is 
T-dual to the conifold, the noncompact Gepner model 
approach \cite{noncompactGepner} might shed light on this issue.

The original idea of identifying the observed fermions as 
coming from 
a ``preon" theory
suffered from the following problems: First, now that the Higgs boson seems to 
have been finally found, there is no evidence for the composite model.
Secondly, the origins of the gauge and gravitational interactions are unclear.
And finally, Kugo-Yanagida's $E_7$ model and its $E_8$ generalization with an 
unbroken $SU(5)$ are anomalous.
One of the virtues of the ``quasi-Nambu-Goldstone fermion 
hypothesis in string theory" proposed in this paper is that it can 
solve, or at least give a possible scenario to solve, all of these 
problems.
Indeed, since we use a symmetry breaking due to branes, the preon theory 
is not needed any more. Gauge interactions and gravity are, of course, 
built in heterotic string theory. Also, the anomaly may be basically canceled by 
an anomaly inflow from the bulk.

This last point requires further discussion. The anomaly polynomial 
for a chiral fermion in four-dimensional Yang-Mills theory is 
$\frac1{48\pi^3}{\rm Tr}F^3$. At first sight, the ${\rm Tr}F^4$ term in $X_8$ 
in the Green-Schwarz counter term may appear to yield the requisite 
contribution if ${\rm Tr}F^4$ 
contains a term of the form ${\rm Tr}F^3 \wedge F_{U(1)}$ in a 
decomposition in terms of some subgroup of $E_8$ and the $F_{U(1)}$ 
develops an expectation value in the transverse space.
However, this is not the case because ${\rm Tr}F^4\propto ({\rm Tr}F^2)^2$ for $E_8$. 
This means that one requires yet another gauge-variant term 
besides the ordinary Green-Schwarz counter term in order to cancel the $SU(5)$ 
anomaly by the anomaly inflow. Although such a correction 
of the supergravity background itself is not surprising in the $(2,0)$ 
worldsheet SUSY models \cite{GvdVZ,DSWW}, 
one would need an explicit form of the field configuration (and not just 
the existence of it) in order to 
 confirm the cancellation by a direct computation.

There are many questions to be asked: How is SUSY broken 
and mediated? How is the GUT group broken? What is the origin 
of the Yukawa hierarchy? How are the dilaton and other moduli 
stabilized? 
Also, although likely to exist,  we haven't so far found an explicit 5-brane field 
configuration that realizes three families of fermions. 
We emphasize, 
however,  
that  
the ``quasi-Nambu-Goldstone fermion 
hypothesis in string theory" proposed in this paper 
is unprecedented in that it might 
at least 
provide a scenario 
{\em explaining why there are three} (or rather, not larger than 
four or five),  
without resorting to the help of the anthropic principle. 
In view of the 
fact that there seems no other viable explanation for it, we believe 
our new framework will deserve further investigations.

\appendix
  
\section{Generators of $E_{8}$ and its subalgebras}
In this appendix we give some detail about $E_8$, in particular 
in terms of Freudenthal's realization, which uses the decomposition 
of $E_8$ into representations of the maximal subgroup $SL(9)$. 
Although it is more familiar to realize $E_8$ as a direct sum of the 
$SO(16)$ adjoint and spinor representations, Freudenthal's realization 
has an advantage in that the generators have only the $SL(9)$ vector 
indices.  It was used by Iri\'{e} and Yasui \cite{IY} to describe 
the K\"ahler coset $E_8/(SO(10)\times SU(3)\times U(1))$
\footnote{Incidentally, it is particularly useful to describe $E_8$ U-duality in three 
dimensions as the $SL(9)$ third-rank tensors correspond to the 
M theory 3-form \cite{E10,MS}.}.

\subsection{$E_{8(+8)}$}
\beqa
\begin{array}{lll}
E^I_{~J} & (I,J=1,\ldots,9; ~~I\neq J) & \mbox{(total 72)}\\
E^{IJK} &(I,J,K=1,\ldots,9)&\mbox{(total 84)}\\
E^*_{IJK} &(I,J,K=1,\ldots,9)&\mbox{(total 84)}\\
h_I &(I=1,\ldots,8)~~(=E^I_{~I}-E^J_{~J})&\mbox{(total 8)}
\end{array}
\eeqa
\beqa
h_{IJK}&\equiv&E^I_{~I}+E^J_{~J}+E^K_{~K}-\frac13\sum_{L=1}^9 E^L_{~L}
\eeqa

\beqa
\begin{array}{lcl}
{[}E^I_{~J},~~E^K_{~L}{]}&=&\delta^{K}_{J} E^I_{~L} -\delta^{I}_{L} E^K_{~J},\\
{[}E^I_{~J},~~E^{KLM}{]}&=&3\delta^{[M}_{I}E^{KL]I},\\
{[}E^I_{~J},~~E^*_{KLM}{]}&=&-3\delta^{I}_{[M}E^*_{KL]J},\\
{[}E^{IJK},~~E^{LMN}{]}&=&-\frac1{3!}\sum_{P,Q,R=1}^9 \epsilon^{IJKLMNPQR}E^*_{PQR},\\
{[}E^*_{IJK},~~E^*_{LMN}{]}&=&+\frac1{3!}\sum_{P,Q,R=1}^9 \epsilon_{IJKLMNPQR}E^{PQR},\\
{[}E^{IJK},~~E^*_{LMN}{]}&=&6\delta^J_{[M}\delta_N^K E^I_{~L]}~~~(\mbox{if $I\neq L,M,N$}),\\
{[}E^{IJK},~~E^*_{IJK}{]}&=&h_{IJK}.
\end{array}
\eeqa

\beqa
{\rm Tr}_{\bf 248}E^I_{~J}E^K_{~L}&=&60\delta^I_L \delta^K_J,\n
{\rm Tr}_{\bf 248}E^{IJK}E^*_{LMN}&=&60\cdot 6 \delta^I_{[L} \delta^J_M \delta^K_{N]}.
\eeqa

\subsection{The compact $E_8$ ($=E_{8(-248)}$)}
\beqa
\begin{array}{lcll}
X_{IJ}&\equiv& E^I_{~J} + E^J_{~I} & (1 \leq I < J \leq 9)\\
Y_{IJ}&\equiv&-i( E^I_{~J} - E^J_{~I}) & (1 \leq I < J \leq 9)\\
h_I&\equiv & E^I_{~I} - E^{I+1}_{~~~I+1} & (I=1,\ldots,8) \\
X_{IJK} &\equiv & E^{IJK} + E^*_{IJK} & (1 \leq I < J< K \leq 9)\\
Y_{IJK} &\equiv & -i(E^{IJK} - E^*_{IJK}) & (1 \leq I < J< K \leq 9)\\
\end{array}
\eeqa

\beqa
\begin{array}{lcl}
{[}X_{IJ},~~X_{KL}{]}&=&+i(\delta_{JK}Y_{IL}+\delta_{IL}Y_{JK}+\delta_{IK}Y_{JL}+\delta_{JL}Y_{IK} ),\\
{[}Y_{IJ},~~Y_{KL}{]}&=&-i(\delta_{JK}Y_{IL}+\delta_{IL}Y_{JK}-\delta_{IK}Y_{JL}-\delta_{JL}Y_{IK} ),\\
{[}X_{IJ},~~Y_{KL}{]}&=&-i(\delta_{JK}X_{IL}-\delta_{IL}X_{JK}+\delta_{IK}X_{JL}-\delta_{JL}X_{IK} ),\\
{[}Y_{IJ},~~X_{KL}{]}&=&-i(\delta_{JK}X_{IL}-\delta_{IL}X_{JK}-\delta_{IK}X_{JL}+\delta_{JL}X_{IK} ),\\
{[}X_{IJ},~~X_{KLM}{]}&=&+3i(\delta_{J[M}Y_{KL]I} +\delta_{I[M}Y_{KL]J}),\\
{[}Y_{IJ},~~X_{KLM}{]}&=&-3i(\delta_{J[M}X_{KL]I} -\delta_{I[M}X_{KL]J}),\\
{[}X_{IJ},~~Y_{KLM}{]}&=&-3i(\delta_{J[M}X_{KL]I} +\delta_{I[M}X_{KL]J}),\\
{[}Y_{IJ},~~Y_{KLM}{]}&=&-3i(\delta_{J[M}Y_{KL]I} -\delta_{I[M}Y_{KL]J}),\\
{[}X_{IJK},~~X_{LMN}{]}&=&+\frac i6 \sum_{P,Q,R=1}^9
\epsilon_{IJKLMNPQR}Y_{PQR}
+18 i \delta_{JM}\delta_{KN}Y_{IL}
~~~{}_{([IJK],[LMN])},\\
{[}Y_{IJK},~~Y_{LMN}{]}&=&-\frac i6 \sum_{P,Q,R=1}^9
\epsilon_{IJKLMNPQR}Y_{PQR}
+18 i \delta_{JM}\delta_{KN}Y_{IL}
~~~{}_{([IJK],[LMN])},\\
{[}X_{IJK},~~Y_{LMN}{]}&=&+\frac i6 \sum_{P,Q,R=1}^9
\epsilon_{IJKLMNPQR}X_{PQR}
+18 i \delta_{JM}\delta_{KN}X_{IL}\\
&&
-2i \delta_{JM}\delta_{KN}\delta_{IL}\sum_{P=1}^9 X_{PP}
~~~{}_{([IJK],[LMN])},\\
\end{array}
\eeqa

In the last line, the trace part of $X$ is projected out, and 
$X_{II} - X_{I+1~ I+1}$ are identified as $2 h_I$.
The symbol ${}_{([IJK],[LMN])}$ means that the rhs must be
anti-symmetrized in the indicated fashion.

\subsection{$E_8 \supset E_7 \times SU(2)$}
\beqa
{\bf 248}&=&({\bf 133},{\bf 1})\oplus ({\bf 1},{\bf 3})\oplus ({\bf 56},{\bf 2})
\nonumber
\eeqa
In the following decompositions, the 
denominator subalgebras are compact ones, whereas 
the coset generators are expressed in the $E_{{8(+8)}}$ form. 
\begin{itemize}
\item{$({\bf 133},{\bf 1})$ $(=E_7)$}
\beqa
\begin{array}{lll}
X_{{\hat i}{\hat j}} & (1\leq {\hat i} < {\hat j} \leq 7) & \mbox{(total 21)}\\
Y_{{\hat i}{\hat j}} & (1\leq {\hat i} < {\hat j} \leq 7) & \mbox{(total 21)}\\
h_{{\hat i}89} &({\hat i}=1,\ldots,7)& \mbox{(total 7)}\\
X_{{\hat i}89}&({\hat i}=1,\ldots,7)& \mbox{(total 7)}\\
Y_{{\hat i}89}&({\hat i}=1,\ldots,7)& \mbox{(total 7)}\\
\\
X_{{\hat i}{\hat j}{\hat k}} &(1\leq {\hat i} < {\hat j} < {\hat k} \leq 7)&\mbox{(total 35)}\\
Y_{{\hat i}{\hat j}{\hat k}} &(1\leq {\hat i} < {\hat j} < {\hat k} \leq 7)&\mbox{(total 35)}\\
\end{array}
\eeqa
Th first five sets  $\{ X_{{\hat i}{\hat j}}, Y_{{\hat i}{\hat j}}, h_{{\hat i}89}, X_{{\hat i}{\hat j}{\hat k}} ,Y_{{\hat i}{\hat j}{\hat k}}   \}$ generate $SU(8)$, while the rest
$\{ X_{{\hat i}{\hat j}{\hat k}} , Y_{{\hat i}{\hat j}{\hat k}}  \}$ form a rank-4 
anti-symmetric tensor representation of $SU(8)$ (cf. \cite{CJ}).

\item{$({\bf 1},{\bf 3})$ $(=SU(2))$ 
}
\beqa
X_{89},~~~Y_{89},~~~h_8~(=E^8_{~8}-E^9_{~9})
\eeqa

\item{$({\bf 56},{\bf 2})$  
}
\beqa
\begin{array}{lll}
E^{\hat i}_{~\alpha} & ({\hat i}=1,\ldots,7;~~\alpha=8,9)& \mbox{(total 14)}\\
E^{\alpha}_{~\hat i} & ({\hat i}=1,\ldots,7;~~\alpha=8,9)& \mbox{(total 14)}\\
E^{{\hat i}{\hat j}\alpha} &(1\leq {\hat i} < {\hat j} \leq 7;~~\alpha=8,9)&\mbox{(total 42)}\\
E^*_{{\hat i}{\hat j}\alpha} &(1\leq {\hat i} < {\hat j} \leq 7;~~\alpha=8,9)&\mbox{(total 42)}\\
\end{array}
\eeqa
\end{itemize}

\subsection{$E_8 (\supset E_7 \times SU(2) )~~~
\supset (SU(5) \times SU(3) \times U(1)) \times SU(2)$}

\begin{center}
\begin{tabular}{c||c|c|c|c|c|c|c|}
Generators 
&\multicolumn{6}{c|}{$E^{\hat i}_{~\alpha}$ $E^{\alpha}_{~~\hat i}$
$E^{{\hat i}{\hat j}\alpha}$ $E^*_{{\hat i}{\hat j}\alpha}$}
&$E^8_{~9}$ $E^9_{~8}$ $h_8$
\\
\renewcommand{\arraystretch}{0.4}
$\begin{array}{c}
E_7\times SU(2) \\
\mbox{representations}
\end{array}$ &\multicolumn{6}{c|}{$({\bf 56},{\bf 2})$}
&$({\bf 1},{\bf 3})$
\\
Generators
&
$E^{i a\alpha}$  $E^a_{~\alpha}$ &
$E^*_{i a\alpha}$  $E^\alpha_{~a}$ &
$E^*_{ab\alpha}$ & 
 $E^{ab\alpha}$ &
$E^{i j\alpha}$  $E^i_{~\alpha}$ &
$E^*_{i j\alpha}$  $E^\alpha_{~i}$ 
&$E^8_{~9}$ $E^9_{~8}$ $h_8$
 \\
\renewcommand{\arraystretch}{0.4}
$\begin{array}{c}
SU(5)\!\times\!SU(3)\!\times\!SU(2)\\
\mbox{representations}
\end{array}$ 
& 
$({\bf 5},{\bf 3},{\bf 2})$&$({\bf \bar 5},{\bf  \bar 3},{\bf 2})$& 
 $({\bf 1},{\bf 3},{\bf 2})$&$({\bf 1},{\bf  \bar 3},{\bf 2})$&
$({\bf 10},{\bf  1},{\bf 2})$& $({\bf \overline{10}},{\bf  1},{\bf 2})$ 
&$({\bf 1},{\bf 1},{\bf 3})$\\
$ h_\sharp$ charge  &
$1$ &
$-1$ &
$-5$ &
$5$ &
$-3$&
$3$ &0
\end{tabular}

\end{center}

\vskip 5ex
(Cont'd)
\begin{center}
\begin{tabular}{|c|c|c|c|c|c|c|c|c|
}
\multicolumn{9}{|c|}{
$E^{\hat i}_{~\hat j}\mbox{\scriptsize$(\hat{i}\neq \hat{j})$}
$
 $E^{{\hat i}89}$ $E^*_{{\hat i}89}$ $h_{{\hat i}89}$ $E^{{\hat i}{\hat j}{\hat k}}$ $E^*_{{\hat i}{\hat j}{\hat k}}$
 }
\\
 \multicolumn{9}{|c|}{$({\bf 133},{\bf 1})$}
\\
$E^{i}_{~j}
$ $E^{i89}$ $E^*_{i89}$ $h_{i89}$ 
&$E^{a}_{~b}
$ $h_{\bar a}$&
$h_\sharp$ &
$E^i_{~a}$ $E^*_{a89}$&
$E^a_{~i}$ $E^{a89}$
& 
$E^*_{ijk}$ $E^{567}$
&$E^{ijk}$ $E^*_{567}$
&$E^*_{ija}$ $E^{iab}$&
$E^{ija}$ $E^*_{iab}$
 \\

$({\bf 24},\! {\bf 1},\! {\bf 1})$& 
$({\bf 1},\!{\bf 3},\!{\bf 1})$&
$({\bf 1},\!{\bf 1},\!{\bf 1})$& 
 $({\bf 5},\!{\bf \bar 3},\!{\bf 1})$&
 $({\bf \bar 5},\!{\bf 3},\!{\bf 1})$&
 $({\bf 5},\!{\bf 1},\!{\bf 1})$& 
$({\bf \bar 5},\!{\bf 1},\!{\bf 1})$&
$({\bf 10},\!{\bf \bar 3},\!{\bf 1})$& 
 $({\bf \overline{10}},\!{\bf 3},\!{\bf 1})$\\
$0$&
$0$& 
$0$& 
$-4$& 
$4$& 
$6$& 
$-6$& 
$2$&
$-2$ 
 \end{tabular}

\end{center}
\beqa
h_\sharp&=&-2(E^1_{~1}+\cdots+E^4_{~4})
+2(E^5_{~5}+E^6_{~6}+E^7_{~7})
+E^8_{~8}+E^9_{~9}.
\label{h_sharp}
\eeqa
In {\em this} table, the values that the various indices run over are:
${\hat i},{\hat j},{\hat k}=1,\ldots,7$; $i,j,k=1,...,4$; $a,b=5,6,7$; ${\bar a}=5,6$.

\subsection{$E_8 \supset E_6 \times SU(3)$}
\beqa
{\bf 248}&=&({\bf 78},{\bf 1})\oplus ({\bf 1},{\bf 8})\oplus ({\bf 27},{\bf 3})
\oplus ({\bf \overline{27}},{\bf \bar 3})
\nonumber
\eeqa
\begin{itemize}
\item{$({\bf 78},{\bf 1})$ $(=E_6)$}
\beqa
\begin{array}{lll}
X_{ij} & (1\leq i < j \leq 6) & \mbox{(total 15)}\\
Y_{ij} & (1\leq i < j \leq 6) & \mbox{(total 15)}\\
h_{\bar i} &({\bar i}=1,\ldots,5)& \mbox{(total 5)}\\
\\
X_{789}&& \mbox{(total 1)}\\
Y_{789}&& \mbox{(total 1)}\\
h_{789}&& \mbox{(total 1)}\\
\\
X_{ijk} &(1\leq i < j <k \leq 6)&\mbox{(total 20)}\\
Y_{ijk} &(1\leq i < j <k \leq 6)&\mbox{(total 20)}\\
\end{array}
\eeqa

\item{$({\bf 1},{\bf 8})$ $(=SU(3))$}
\beqa
\begin{array}{lll}
X_{ab} & (7\leq a < b \leq 9) & \mbox{(total 3)}\\
Y_{ab} & (7\leq a < b \leq 9) & \mbox{(total 3)}\\
h_{\bar a} &({\bar a}=7,8)& \mbox{(total 2)}\\
\end{array}
\label{SU(3)subalgebra}
\eeqa

\item{$({\bf 27},{\bf 3})$}
\beqa
\begin{array}{lll}
E^a_{~i} & (i=1,\ldots,6;~a=7,8,9) & \mbox{(total $6\times 3$)}\\
E^*_{iab} & (i=1,\ldots,6;~7\leq a<b \leq 9) & \mbox{(total $6\times 3$)}\\
E^{ija} &(1\leq i < j \leq 6;~a=7,8,9)& \mbox{(total $15\times 3$)}\\
\end{array}
\eeqa

\item{$({\bf \overline{27}},{\bf \bar3})$}
\beqa
\begin{array}{lll}
E^i_{~a} & (i=1,\ldots,6;~a=7,8,9) & \mbox{(total $6\times 3$)}\\
E^{iab} & (i=1,\ldots,6;~7\leq a<b \leq 9) & \mbox{(total $6\times 3$)}\\
E^*_{ija} &(1\leq i < j \leq 6;~a=7,8,9)& \mbox{(total $15\times 3$)}\\
\end{array}
\eeqa

\end{itemize}

\subsection{$E_8  (\supset E_6 \times SU(3) )~~~
\supset (SO(10)  \times U(1)) \times SU(3) \\
~~~~~~~~~~~~~~~~~~~~~~~~~~~~~~~~~
\supset (SU(5) \times U(1) \times U(1)) \times SU(3)$}

\begin{center}
\begin{tabular}{c||c|c|c|c|c|c|c|c|c|c|c|}
Generators 
&\multicolumn{11}{c|}{
$E^{i}_{~j}(i\neq j)$ 
$h_{\bar i}$
$E^{ijk}$ $E^*_{ijk}$ $E^{789}$ $E^*_{789}$}
\\
\renewcommand{\arraystretch}{0.4}
$\begin{array}{c}
E_6\times SU(3) \\
\mbox{representations}
\end{array}$ &\multicolumn{11}{c|}{$({\bf 78},{\bf 1})$}
\\
Generators
&$E^{\bar i}_{~\bar j}$ $h_{\bar{\bar i}}$
&$E^{\bar i \bar j 6}$ 
&$E^*_{\bar i \bar j 6}$
&$h_{456}$
&$E^*_{\bar i \bar j \bar k}$ 
&$E^6_{~\bar i}$
&$E^{789}$
&$E^{\bar i \bar j \bar k}$ 
&$E^{\bar i}_{~6}$
&$E^*_{789}$
&$h_\perp$
 \\
\renewcommand{\arraystretch}{0.4}
$\begin{array}{c}
SU(5)\times SU(3) \\
\mbox{representations}
\end{array}$ 
& $({\bf 24},\!{\bf 1})$
& $({\bf 10},\!{\bf 1})$
& $({\bf \overline{10}},\!{\bf 1})$
& $({\bf 1},\!{\bf 1})$
& $({\bf 10},\!{\bf 1})$
& $({\bf\bar 5},\!{\bf 1})$
& $({\bf 1},\!{\bf 1})$
& $({\bf \overline{10}},\!{\bf 1})$
& $({\bf 5},\!{\bf 1})$
& $({\bf 1},\!{\bf 1})$
& $({\bf 1},\!{\bf 1})$

\\
\renewcommand{\arraystretch}{0.4}
$\begin{array}{c}
SO(10)\times SU(3) \\
\mbox{representations}
\end{array}$ 
& \multicolumn{4}{c|}{$({\bf 45},\!{\bf 1})$
}
&  \multicolumn{3}{c|}{$({\bf 16},\!{\bf 1})$
}
&  \multicolumn{3}{c|}{$({\bf \overline{16}},\!{\bf 1})$
}
& $({\bf 1},\!{\bf 1})$

\\
$3h_\perp$ charge
&
\multicolumn{4}{c|}{$0$}
&
\multicolumn{3}{c|}{$3$}
&
\multicolumn{3}{c|}{$-3$}
&$0$
\\
$3h'_\perp$ charge
&
$0$
&$3$
&$-3$
&$0$
&$0$
&$3$
&$-3$
&$0$
&$-3$
&$3$
&$0$

\end{tabular}

\end{center}


(Cont'd)
\begin{center}

\begin{tabular}{|c|c|c|c|c|c|c|c|c|c|c|c|c|}
\multicolumn{6}{|c|}{$E^a_{~i}$ $E^*_{iab}$ $E^{ija}$}
&\multicolumn{6}{c|}{$E^i_{~a}$ $E^{iab}$ $E^*_{ija}$}
&$E^a_{~b}$ $h_{\bar a}$
\\
\multicolumn{6}{|c|}{$({\bf 27},{\bf 3})$}
&\multicolumn{6}{c|}{$({\bf \overline{27}},{\bf \bar 3})$}
&$({\bf 1},{\bf 8})$
\\
$E^{\bar i \bar j a}$
& $E^*_{\bar i a b}$
&$E^a_{~6}$
&$E^{\bar i 6 a}$
&$E^a_{~\bar i}$
&$E^*_{6ab}$
&$E^*_{\bar i \bar j a}$
& $E^{\bar i a b}$
&$E^6_{~a}$
&$E^*_{\bar i 6 a}$
&$E^{\bar i}_{~a}$
&$E^{6ab}$
&$E^a_{~b}$ $h_{\bar a}$
\\
$({\bf 10},\!{\bf 3})$
&$({\bf \bar 5},\!{\bf 3})$
&$({\bf 1},\!{\bf 3})$
&$({\bf 5},\!{\bf 3})$
&$({\bf \bar 5},\!{\bf 3})$
&$({\bf 1},\!{\bf 3})$
&$({\bf \overline{10}},\!{\bf 3})$
&$({\bf 5},\!{\bf 3})$
&$({\bf 1},\!{\bf 3})$
&$({\bf \bar 5},\!{\bf 3})$
&$({\bf 5},\!{\bf 3})$
&$({\bf 1},\!{\bf 3})$
&$({\bf 1},\!{\bf 8})$
\\
 \multicolumn{3}{|c|}{$({\bf 16},{\bf 3})$}
 & \multicolumn{2}{c|}{$({\bf 10},{\bf 3})$}
 &$({\bf 1},{\bf 3})$
 &\multicolumn{3}{c|}{$({\bf \overline{16}},{\bf 3})$}
 & \multicolumn{2}{c|}{$({\bf 10},{\bf 3})$}
 &$({\bf 1},\!{\bf 3})$
 &$({\bf 1},\!{\bf 8})$
 \\
  \multicolumn{3}{|c|}{$-1$}
 & \multicolumn{2}{c|}{$2$}
 &$-4$
 &\multicolumn{3}{c|}{$1$}
 & \multicolumn{2}{c|}{$-2$}
 &$4$
 &$0$
 \\
$-1$
&$2$
&$-4$
&$2$
&$-1$
&$-1$
&$1$
&$-2$
&$4$
&$-2$
&$1$
&$1$
&$0$
\end{tabular}

\end{center}


\beqa
h_{\perp}&=&\frac13\left(
-(E^1_{~1}+\cdots E^5_{~5})+2 E^6_{~6}+(E^7_{~7}+E^8_{~8}+E^9_{~9})
\right).
\label{h_perp}
\eeqa

\beqa
 h'_\perp&=&E^6_{~6}-\frac13(E^7_{~7}+E^8_{~8}+E^9_{~9}).
 \label{h_perp'}
\eeqa

\section{Conventions for the gamma matrices}
\label{conventions}
\beqa
\Gamma^a&=&\gamma^a_{4D} \otimes {\bf 1},\\
\Gamma^\alpha&=&\gamma^\sharp_{4D} \otimes \gamma^\alpha,
\eeqa
where $a=0,7,8,9$ and $\alpha=1,2,3,4,5,6$.
\beqa
{\{} \Gamma^A,~~\Gamma^B  {\}}&=&2\eta^{AB},\\
\eta^{AB}&=&\mbox{diag}(-+\cdots +). \nonumber
\eeqa

\beqa
\gamma^0_{4D}&=&i\sigma_2 \otimes {\bf 1}, \n
\gamma^7_{4D}&=&\sigma_1 \otimes \sigma_1, \n
\gamma^8_{4D}&=&\sigma_1 \otimes \sigma_2, \\
\gamma^9_{4D}&=&\sigma_1 \otimes \sigma_3, \n
\gamma^\sharp_{4D}&=&\sigma_3 \otimes {\bf 1}
~=~-i \gamma^0_{4D}\gamma^7_{4D}\gamma^8_{4D}\gamma^9_{4D}. \nonumber
\eeqa

\beqa
\gamma^1&=&\sigma_2\otimes{\bf 1}\otimes{\bf 1},\n
\gamma^2&=&\sigma_1\otimes \sigma_1\otimes{\bf 1},\n
\gamma^3&=&\sigma_1\otimes \sigma_2\otimes{\bf 1},\n
\gamma^4&=&\sigma_1\otimes \sigma_3\otimes \sigma_1,\\
\gamma^5&=&\sigma_1\otimes \sigma_3\otimes \sigma_2,\n
\gamma^6&=&\sigma_1\otimes \sigma_3\otimes \sigma_3,\n
\gamma^\sharp&=&\sigma_3\otimes{\bf 1}\otimes{\bf 1}
~=~-i \gamma^1\gamma^2 \cdots \gamma^6.\nonumber
\eeqa
\beqa
\Gamma_{11}&=&\gamma^\sharp_{4D}\otimes \gamma_\sharp.
\eeqa
\beqa
B&=&\Gamma^8 \Gamma^1 \Gamma^3 \Gamma^5. 
\eeqa

\section*{Acknowledgments} 
We would like to thank T.~Kugo for valuable discussions.
This work was stimulated by remarks made to one of the 
authors (S.M.) some time ago by H.~Kunitomo, whom 
we also acknowledge here. We also thank S.~Aoyama, 
T.~Kimura, N.~Kitazawa, T.~Kobayashi, H.~Kodama, K.~Kohri 
and Y.~Sakamura for discussions.
The work of S.~M. is supported by 
Grant-in-Aid
for Scientific Research  
(A) No.22244030 and (C) No.20540287  
from
The Ministry of Education, Culture, Sports, Science
and Technology of Japan.
The work of M.~Y. is supported by 
JSPS.

\end{document}